\documentclass[sigconf]{acmart}
\usepackage{graphicx}
\usepackage{subcaption}
\usepackage{multirow}
\usepackage{multicol}
\AtBeginDocument{%
  \providecommand\BibTeX{{%
    \normalfont B\kern-0.5em{\scshape i\kern-0.25em b}\kern-0.8em\TeX}}}

\setcopyright{acmlicensed}
\copyrightyear{2018}
\acmYear{2018}
\acmDOI{XXXXXXX.XXXXXXX}

\acmConference[Conference acronym 'XX]{Make sure to enter the correct
  conference title from your rights confirmation emai}{June 03--05,
  2018}{Woodstock, NY}
\acmISBN{978-1-4503-XXXX-X/18/06}

\makeatletter
\def\@authornotemark{}
\makeatother

\begin{document}

\title{Contrastive Learning for Implicit Social Factors in Social Media Popularity Prediction}
\settopmatter{printacmref=false} 
\renewcommand\footnotetextcopyrightpermission[1]{}
\author{Zhizhen Zhang}
\email{zhangzz21@tsinghua.org.cn}
\affiliation{%
  \institution{Tsinghua University}
  \city{Beijing}
  \country{China} 
}
\authornote{This work was completed by Zhizhen Zhang at Tsinghua University, and he has since graduated.}

\author{Ruihong Qiu}
\email{r.qiu@uq.edu.cn}
\affiliation{%
  \institution{The University of Queensland}
  \city{Brisbane}
  \country{Australia}
}

\author{Xiaohui Xie}
\email{xiexiaohui@mail.tsinghua.edu.cn}
\affiliation{%
  \institution{Tsinghua University}
  \city{Beijing}
  \country{China}
}


\begin{abstract}
On social media sharing platforms, some posts are inherently destined for popularity. Therefore, understanding the reasons behind this phenomenon and predicting popularity before post publication holds significant practical value. The previous work predominantly focuses on enhancing post content extraction for better prediction results. However, certain factors introduced by social platforms also impact post popularity, which has not been extensively studied. For instance, users are more likely to engage with posts from individuals they follow, potentially influencing the popularity of these posts. We term these factors, unrelated to the explicit attractiveness of content, as implicit social factors. Through the analysis of users' post browsing behavior (also validated in public datasets), we propose three implicit social factors related to popularity, including content relevance, user influence similarity, and user identity. To model the proposed social factors, we introduce three supervised contrastive learning tasks. For different task objectives and data types, we assign them to different encoders and control their gradient flows to achieve joint optimization. We also design corresponding sampling and augmentation algorithms to improve the effectiveness of contrastive learning. Extensive experiments on the Social Media Popularity Dataset validate the superiority of our proposed method and also confirm the important role of implicit social factors in popularity prediction. We open source the code at https://github.com/Daisy-zzz/PPCL.git.
\end{abstract}



\maketitle

\section{Introduction}
As social networks become increasingly popular, more and more users are joining these platforms. Especially on content sharing platforms, users interact extensively by creating various posts including images, titles, personalized tags, etc. An intriguing observation is that some posts are born to be popular, making predicting the popularity of these posts before their publication highly valuable in realms such as online advertising and recommendation ~\cite{pinto2013using,khosla2014makes,he2014predicting}. \par 
 
Appealing content can contribute to the popularity of a post, such as engaging photos and interesting captions. Thus, previous studies have concentrated on enhancing the representation of the post content, e.g., images and captions ~\cite{lai2020hyfea, wu2022deeply, chen2023}. Many of these approaches require manual extraction of a variety of image and text features, and the extensive feature engineering leads to increased complexity, time-consuming processes, and hinders the practical application of these methods ~\cite{wang2017combining,he2019feature,kang2019catboost,xu2020multimodal, 
lai2020hyfea, wu2022deeply,Hsu2023}. Some recent methods have adopted pre-trained models to extract content features, subsequently fine-tuning them with downstream tasks to enhance the quality of post content representations in an end-to-end manner ~\cite{tan2022efficient,chen2022and,WANG2023101490,LIU2023103738,chen2023}. Additionally, some others utilize temporal models to aggregate historical post content sequences, aiming to improve the prediction of current post popularity ~\cite{wu2017sequential,wang2020feature, tan2022efficient,zhang2023improving}.\par


While the appeal of the content holds significant importance in determining the popularity of a post, various other factors within social platforms contribute to popularity as well. Previous study on social marketing theories ~\cite{doi:10.1016/j.intmar.2020.05.001,doi:10.1177/1783591719847545} shows that the content format and platform directly affect users’ passive and active engagement behavior. For instance, some users are more inclined to browse posts that are recommended to them, and some others prefer to click on posts from people they follow, rather than posts that are inherently appealing in content. We categorize these platform-induced elements as implicit social factors, which are distinct from the explicit attractiveness of the content. These factors will subtly affect the behavior patterns of users browsing posts, and thus affect the popularity of posts. According to the influence of the content and platform, we first conduct a qualitative analysis of affected users' viewing post behavior and propose the following hypothesis: (1) Users tend to browse posts that align with their content preferences. Thus, posts with similar content are likely to have overlapping audiences, leading to their correlated popularity. (2) Users also tend to browse posts published by users they follow. Therefore, for a particular publisher, the popularity of his posts is more relevant with a close fan base. (3) In the case of different publishers, the size of user influence can impact the exposure of their posts, i.e., the more followers a user has, the greater the chances of his posts being seen. Consequently, posts from users with similar influence might exhibit closer levels of popularity. We summarize the above factors as \textbf{Content Relevance (CR)}, \textbf{User Identity (UI)}, and \textbf{User Influence Similarity (UIS)}.  \par

\begin{figure*}[t]

  \begin{subfigure}{0.36\textwidth}
    \includegraphics[width=\linewidth]{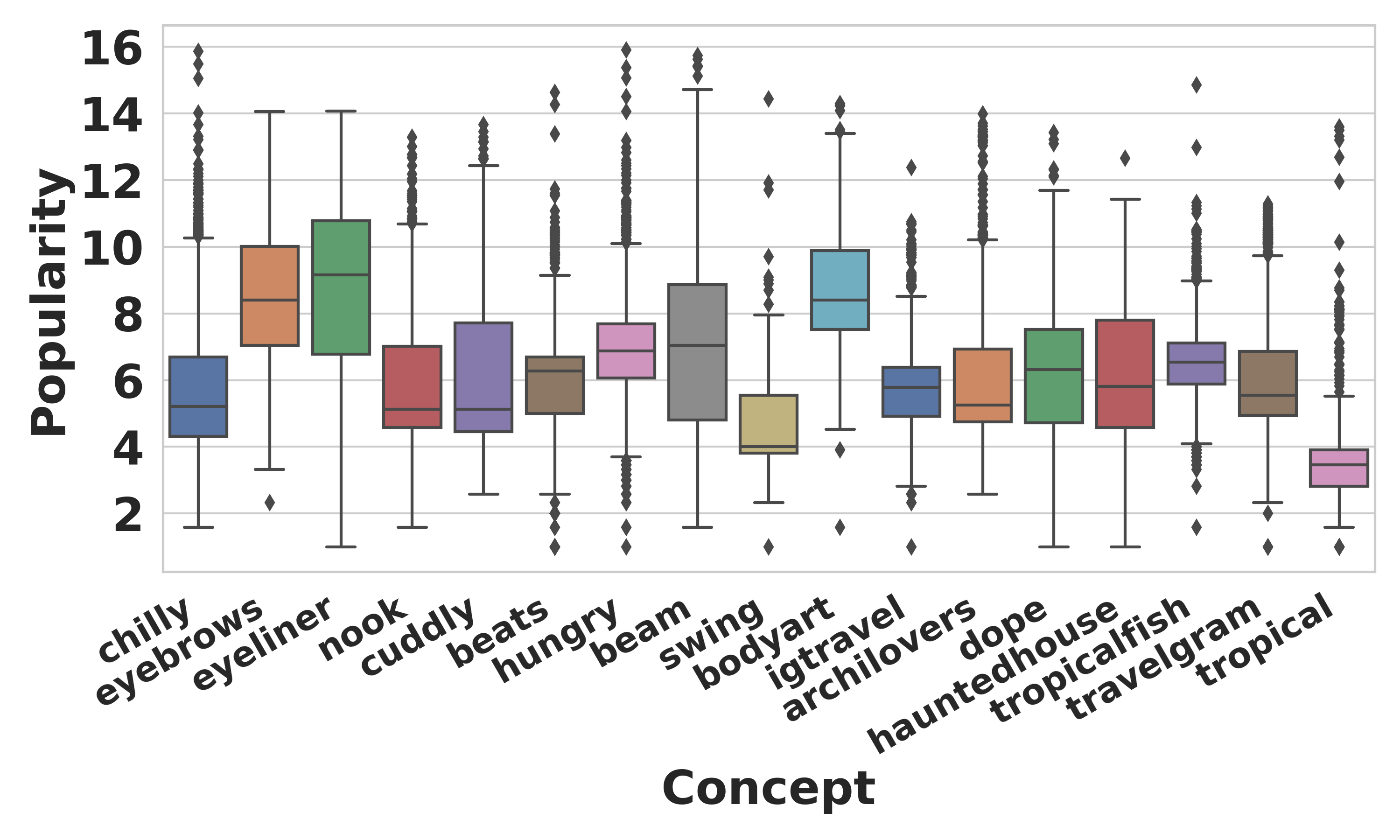}
    \caption{Box plot of popularity distribution across different categories}
    \label{fig:motistats-sub2}
  \end{subfigure}
  \hfill
  \begin{subfigure}{0.26\textwidth}
    \includegraphics[width=\linewidth]{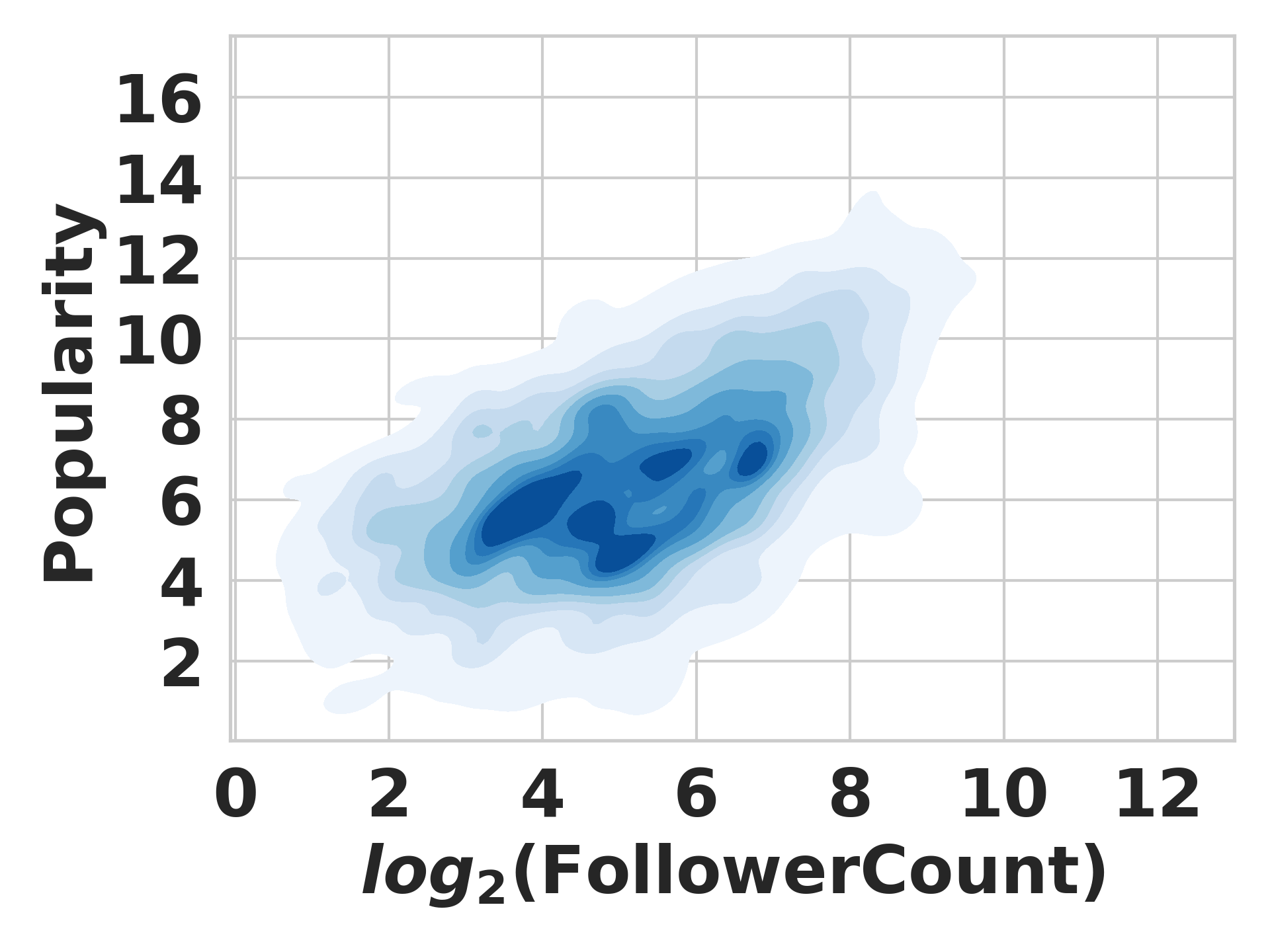}
    \caption{Density plot of popularity vs. log-normalized FollowerCount}
    \label{fig:motistats-sub1}
  \end{subfigure}
  \hfill
  \begin{subfigure}{0.36\textwidth}
    \includegraphics[width=\linewidth]{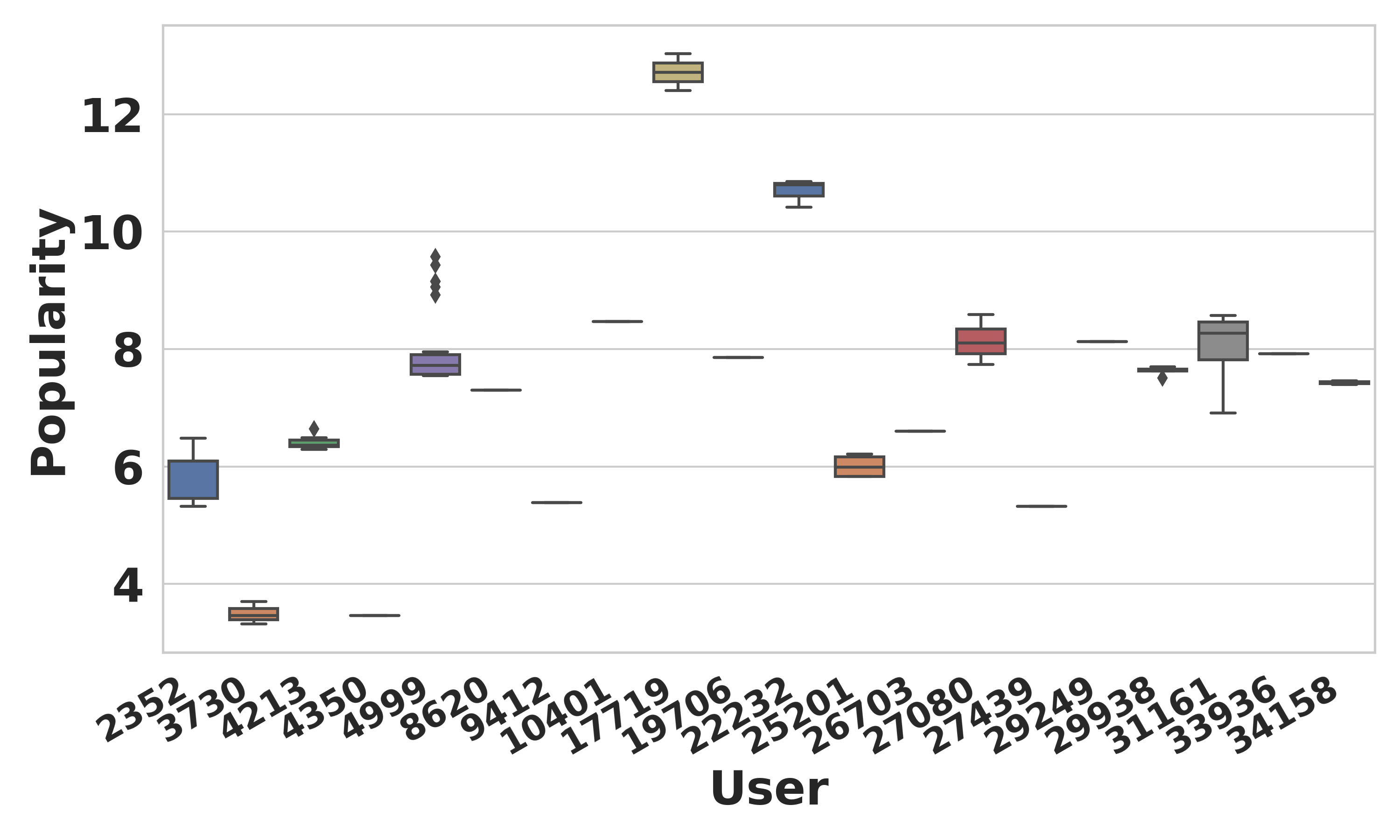}
    \caption{Box plot of popularity distribution across different users}
    \label{fig:motistats-sub3}
  \end{subfigure}
  
  \caption{Popularity distributions across three data attributes in SMPD, respectively validating the impact of proposed social factors on popularity: Content Relevance (CR), User Influence Similarity (UIS), and User Identity (UI).}
  \label{fig:motistats}
\end{figure*}

To verify the rationality of the motivation, we perform a quantitative analysis of the widely used Social Media Popularity Dataset (SMPD) ~\cite{wu2019smp}, which includes information about posts published on Flickr. In Figure ~\ref{fig:motistats-sub2}, we randomly select 20 Concepts and corresponding posts. Different Concept labels in SMPD denote different content categories of posts. Since Concept is categorical, we present box plots of the popularity distribution of each Concept, revealing obvious variations in the popularity of posts across different content categories and relative concentration of popularity within the same category, which validates factor CR. In Figure ~\ref{fig:motistats-sub1}, we choose the FollowerCount attribute to denote the user influence. Since the FollowerCount of different users is continuous real numbers varying greatly, we draw the 2D density plot where x-axis denotes $log_2$(FollowerCount) and y-axis denotes popularity. As we can see, the density reveals a more narrow distribution along the y-axis, indicating that users with similar influence tend to have posts that are relatively close in popularity, which validates factor UIS. In Figure ~\ref{fig:motistats-sub3}, we randomly choose 20 users and their posts, the x-axis denotes the selected userID. It can be found that there are significant differences between the popularity distribution of posts by different users, and the high concentration of popularity among posts from the same user, which validates factor UI. These three observations effectively validate our hypothesis on the social factors that influence popularity. Note that this is a preliminary verification of motivation through quantitative analysis. We will later prove the effectiveness of the proposed social factors for popularity prediction through comprehensive experiments.

Notice that some previous works have adopted network-based methods to study the factor like UIS ~\cite{li2017deepcas,cao2017deephawkes}. Their approach is based on cascade graphs of post retweets or relationship networks of users to find the relationship between connected users and posts. However, this explicit interaction information is not available in the scenarios we study. All we have access to is post content and static user data. Therefore, to incorporate social factors without explicit interactive information, we abstract them as the similarity of data in three dimensions: post content, user identity, and user influence. This enlightens us to think of them as contrastive signals, using contrastive learning in the feature space to bring the representations of samples that are similar in each dimension closer while pushing apart those of dissimilar samples. Specifically, we introduce an end-to-end framework PPCL, which consists of a base prediction architecture and additional contrastive learning modules. PPCL includes post and user encoders to encode post content and user information in a common space, alongside a popularity predictor that integrates all information for predictions. To incorporate implicit social factors, PPCL further adds three supervised contrastive learning tasks to jointly optimize different components of the model: (1) Content Relevance Discrimination task. Tailored for the post encoder, this task aims to differentiate between posts with diverse content. This enhances the model's ability to understand post content at a fine-grained level. (2) User Influence Similarity Discrimination task. This task is designed for the user encoder to distinguish users with varying levels of influence. It enables the model to generate user representations that are sensitive to different levels of the publisher's influence. (3) User Identity Discrimination task. Focused on the popularity predictor, this task facilitates consistent predictions for posts originating from the same user. By assigning the three contrastive learning tasks to different encoders and controlling their gradient flows, the representation capabilities of these encoders are improved. We also design a corresponding sampling process as well as an unsupervised sample augmentation method to further enhance the learning of these tasks acting on different data sources and encoders. Further experimental results reveal that the three proposed social factors can serve as inductive biases in the popularity prediction task, enabling the model to easily learn popularity patterns even from limited data.

In summary, our main contributions are as follows:
\begin{itemize}
    \item We introduce three implicit social factors, i.e.,  content relevance, user influence similarity, and user identity, which play roles in determining post popularity.
    \item We design a comprehensive framework PPCL that consists of post and user encoders, along with a popularity predictor, jointly optimized through three supervised contrastive learning tasks targeting the proposed social factors.
    \item We conduct extensive experiments on the Social Media Popularity Dataset under three different settings. The results show the superiority of our proposed method and validate the significant impact of social factors on popularity.

\end{itemize}

\section{related work}

For the Social Media Popularity Prediction task, conventional works adopt multiple feature extractors to extract image and text features of posts, fusing them with other metadata (\textit{e.g.} user information) and making predictions with regression models \cite{chen2019social, ding2019social,xu2020multimodal,wu2022deeply,Hsu2023}. For example, Wu \textit{et al.} \cite{wu2022deeply} adopts CLIP, Nima, HyperIQA, and Place365 to extract image features and use BERT, the text branch of CLIP to extract text feature. Such kinds of methods require extracting features from several heavy pre-trained models for each new post, which is bulky and unpractical in real scenarios. Recently, some researchers have started to design models in an end-to-end manner. Chen \textit{et al.} \cite{chen2022and} build two-stream ViLT models for title-visual and tag-visual representations, and design title-tag contrastive learning for two streams to learn the differences between titles and tags. Tan \textit{et al.} \cite{tan2022efficient} perform visual and textual feature extraction respectively and then employ a multimodal transformer ALBEF to align visual and text features in semantic space. Chen \textit{et al.} \cite{chen2023} pre-train the vision-and-language transformer by multi-task learning, finetuning two models with different training strategies, and ensemble their results. Wang \textit{et al.} \cite{WANG2023101490} adopt a two-stage fusion method to fuse intra-modality and inter-modality features, respectively. Liu \textit{et al.} \cite{LIU2023103738} extract the relationship between objects in the image, enhancing the understanding of the post content. There are also some works that adopt sequence modeling methods to utilize historical posts to improve the prediction of the current post. Wu \textit{et al.} \cite{wu2017sequential} analyze temporal characteristics of social media popularity, consider the posts as temporal sequences, and make a prediction with temporal coherence across multiple time scales. 
Tan \textit{et al.} \cite{tan2022efficient} adopt a time transformer to aggregate the user's historical post content sequence. 
Zhang \textit{et al.} \cite{zhang2023improving} use LSTM and attention mechanisms to aggregate content-related posts in a size-fixed time window to enhance the current post feature. Previous works commonly face a limitation as they heavily depend on post content extraction. They mine the impact of the content itself on popularity through extensive feature engineering or pre-training and fine-tuning of the pre-trained model. However, this approach makes it difficult to uncover implicit social factors in the data that are also important for popularity prediction. 

Some other work uses information cascades to study social media popularity ~\cite{li2017deepcas, cao2017deephawkes, 10.1145/3580305.3599281}. They primarily utilize the cascade graph formed by post reposts or the users' connection networks to model the impact of users or propagation structures on the popularity of posts. In our scenario, we can only utilize post content and static user information, so we do not elaborate further on this aspect.
\section{Problem Definition}
Formally, given a new post $p$ published by user $u$, the problem of predicting its popularity is to estimate how much attention it would receive after its release (\textit{e.g.} views, clicks or likes \textit{etc.}). In Social Media Popularity Dataset \cite{wu2019smp} which we use for the experiment, “viewing count” is used to compute the popularity label $\hat{y}$ as below:
\begin{equation}
\hat{y}=\log _2 \frac{r}{d}+1
\end{equation}
where $r$ is the viewing count, $d$ is the number of days since the photo was posted. In this way, $\hat{y}$ denotes a normalized measure of a post’s popularity that accounts for both the viewing count and the time since the post was made. It provides a more balanced view of how popular a post is relative to the amount of time it has been available, facilitating fairer comparisons between posts with different viewing durations.\par

In this paper, for any given post $p$, we use the image $I$ and text $T$ in that post and its publisher information $U$ to predict its popularity. Our objective is to construct an end-to-end model $\mathcal{M}$ that can generate popularity predictions $y$ given inputs $I,T,U$:
$$
y=\mathcal{M}(I,T,U;\Theta)
$$
where $\Theta$ is the parameter of $\mathcal{M}$. In our design, $\Theta$ is jointly optimized by the regression loss supervised by the popularity label, as well as the contrastive losses designed for different encoders in $\mathcal{M}$. The joint optimization objective can be expressed as:
$$
\text{minimize } \mathcal{L}(\Theta) = \lambda \cdot \mathcal{L}_{\text{reg}}(\Theta) + (1-\lambda)\sum_{i}\alpha_i \cdot \mathcal{L}_{\text{contra}_i}(\Theta_i)
$$
where $\Theta_i \subset \Theta$ and $\Theta_i \cap \Theta_j = \emptyset$ for $i \neq j$. $\lambda$ and $\alpha_i$ are coefficients that control the influence of each loss during optimization.
\section{the proposed model}
\begin{figure*}[t]
    \centering
    \includegraphics[width=0.8\textwidth]{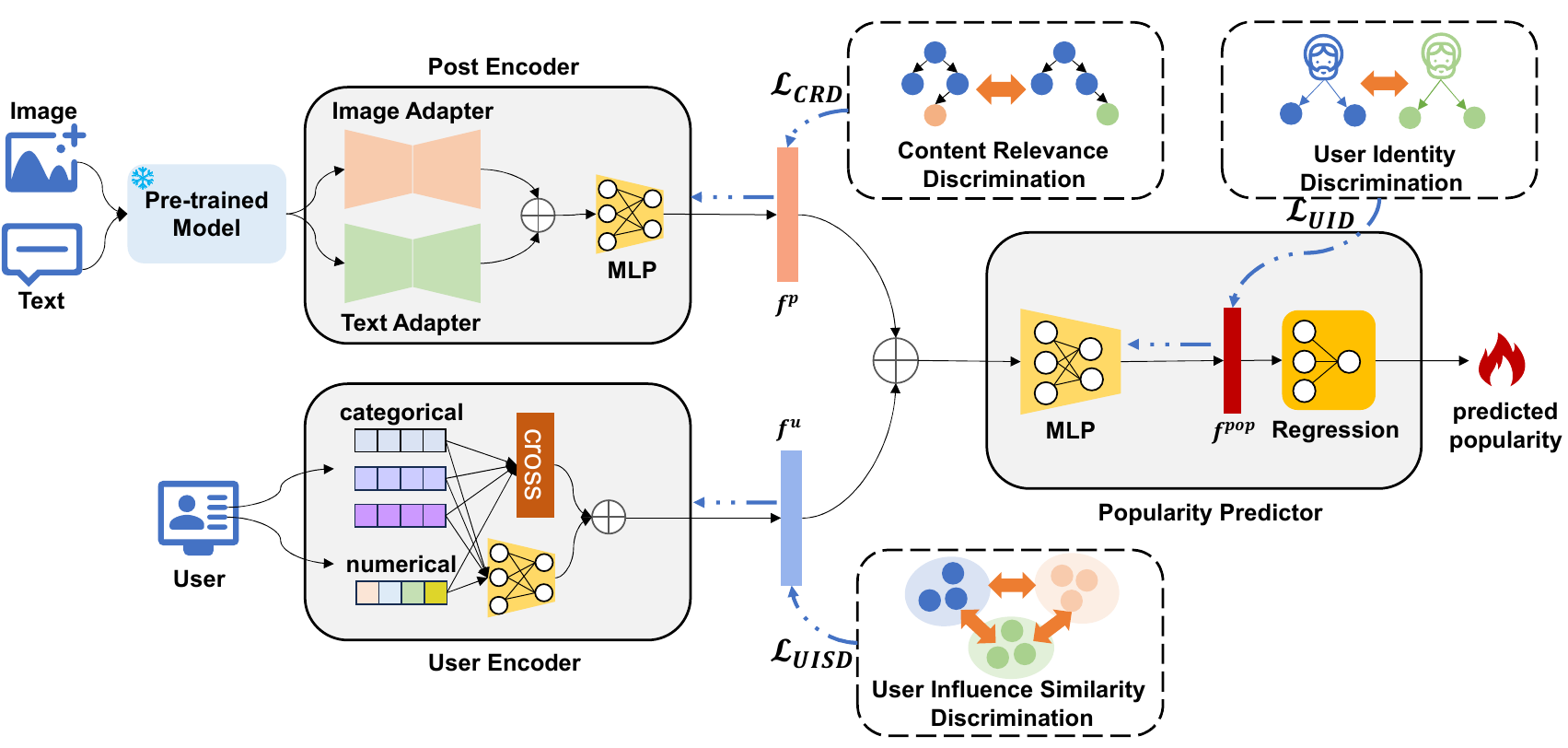}
    \caption{An overview of the proposed PPCL. The model architecture consists of post encoder, user encoder, and popularity predictor. Three contrastive learning losses, i.e., $\mathcal{L}_{CRD}$, $\mathcal{L}_{UISD}$, and $\mathcal{L}_{UID}$ are then added to optimize each of the above components for modeling the proposed implicit social factors. }
    \label{fig:model}
\end{figure*}

We design our model in an end-to-end manner, as shown in Figure~\ref{fig:model}, incorporating (1) Post Encoder that encodes image and text information of the post and learns a multi-modal representation. (2) User Encoder that encodes category and numerical user attributes, and (3) Popularity Predictor that combines all the information to output popularity features and makes final predictions. To capture implicit social influences, we introduce three contrastive learning tasks, jointly optimizing different components. This section will first provide a detailed introduction to the base prediction architecture, the contrastive learning part will be elaborated later.

\subsection{Post Encoder}
A post on Flickr usually consists of an image and some text information, including a post title and several customized hashtags. The text information is a description of the image, so we aim to use multi-modal learning to extract consistent image-text features. Due to the success of pre-trained models in multi-modal learning, we adopt CLIP ~\cite{radford2021learning}, a multi-modal model pre-trained on a variety of (image, text) pairs, to extract raw features. Given the image $I$ of a target post, we construct corresponding text $T$ as ``The title of the image is \{title\} and the tags are \{hashtags\}''. Then the $(I,T)$ pair is input into CLIP:
\begin{align}
    f^{v}_r &= \text{CLIP-image}(I) \\
    f^{t}_r &= \text{CLIP-text}(T)
\end{align}
The image and text branches of CLIP are used to extract the raw image and text features $f^{v}_r, f^{t}_r \in \mathcal{R}^{d_{r}}$, where $d_{r}$ is the output dimension of CLIP. To obtain more task-specific representations, we adopt a lightweight method to adapt raw features to the new dataset. For computational efficiency, we follow ~\cite{gao2021clip} to freeze the backbone of CLIP and use additional learnable bottleneck layers to learn new features:
\begin{align}
    f^{v} &= \text{ReLU}({f^{v}_{r}}^T \cdot W_1^v)\cdot W_2^v \\
    f^{t} &= \text{ReLU}({f^{t}_{r}}^T \cdot W_1^t)\cdot W_2^t 
\end{align}
ReLU is the activation function, $W_1^* \in \mathcal{R}^{d_r \times d_b}$ is the parameters of bottleneck linear layers where $d_b < d_r$, $W_2^* \in \mathcal{R}^{d_b \times d_h}$ where $d_h = d_r$ transform features into the common hidden space. Then we use a two-layer Multi-layer Perception (MLP) to capture the relationship between image and text features:
\begin{equation}\label{eq:pmlp}
    f^p = \text{MLP}(\left[f^v,f^t\right])
\end{equation}
where $\left[\cdot,\cdot\right]$ is concatenation. After fusion, we get $f^p \in \mathcal{R}^{d_h}$ as the multi-modal representation of the post. Note that CLIP can be easily replaced by other lightweight pre-trained models without much performance loss, which we verify in Appendix ~\ref{app:param}.

\subsection{User Encoder}
 The inputs of user information are usually in two kinds of structures: numerical (dense) and categorial (sparse) features. We use $D=\left[D_1,D_2,...,D_N\right]$ to denote dense features including $N$ numerical fields and $S=\left[S_1,S_2,...,S_M\right]$ to denote sparse features including $M$ categorical fields. Since there can be higher-order relationships in user information, we use two approaches, feature crossing which has been widely validated in the domain of click-through rate (CTR) prediction ~\cite{yu2021xcrossnet,wang2023deeper}, and MLP to both explicitly and implicitly capture interactions between different input fields. 

 \subsubsection{\textbf{Feature crossing}}\par
 We first introduce the cross-layers, each of which performs explicit interactions between input features. Then the model captures the high-order relationships by stacking several cross-layers.
\paragraph{\textbf{Cross-layer}}
Given $X_a, X_b \in d_c$ as the input with feature structure $t$ of the $i_{\text{th}}$ cross-layer, we adopt inner-product and linear transformations to perform the crossing process:
\begin{equation}
    \text{Cross}_i^t\left(X_a,X_b\right) = {X_a}^T \cdot X_b \cdot W_i^t + b_i^t
\end{equation}
where $W_i^t, b_i^t \in \mathcal{R}^{d_c}$ are the weight and bias parameters, respectively. We use different parameters to model diverse relationships between different structures of features (i.e., dense or sparse).

\paragraph{\textbf{Dense feature crossing}}
Since each field of the dense feature $D$ is a real number that indicates different user attributes, we denote the dimension of $D$ as $d_N$ which equals the number of numerical fields $N$. Then we adopt $l$ cross-layers to capture $\left(1,2,...,l\right)$-order relationships between each dimension of $D$:
\begin{equation}
    O_i^d=\text{Cross}_i^d\left(D,O_{i-1}^d\right), i=1,...l
\end{equation}
where $O_i^d \in \mathcal{R}^{d_N}$ are the output of each numerical cross-layers, when $i=1$, $O_0^d$ is the original dense feature $D$. We concatenate outputs of $l$ layers to combine different order relationships between dense features:
\begin{equation}
    O^d = \left[O_0^d, O_1^d, ..., O_l^d\right]
\end{equation}
where $O^d \in \mathcal{R}^{(l+1) * d_N}$ is the output of numerical feature crossing.

\paragraph{\textbf{Sparse feature crossing}}
As sparse features $S$ are represented as vectors of one-hot encoding of high-dimensional spaces, we employ an embedding layer to transform these one-hot encoding vectors into dense vectors $E$ as:
\begin{equation}
\begin{aligned}
    E &= \left[E_1,E_2,...,E_M\right] \\
    E_i &= \text{embed}\left(S_i\right), i=1,...,M 
\end{aligned}
\end{equation}
where $S_i$ indicates the input sparse feature of categorical field $i$, $E_i \in \mathcal{R}^{d_M}$ indicates the feature embedding of field $i$ with dimension $d_M$. We adopt $M * \left(M-1\right) / 2$ cross-layers to extract relationships of each embedding pair of $E$:
\begin{equation}
    O_{i,j}^s=\text{Cross}_{i,j}^s\left(E_i,E_j\right),i \neq j
\end{equation}
where $O_{i,j}^s$ is the crossed features between categorical fields $i$ and $j$. We concatenate individual and crossed dense features as the output of sparse feature crossing:
\begin{equation}
    O^s=\left[E_1,...,E_M,O_{1,2}^s,...,O_{M-1,M}^s\right]
\end{equation}
The dimension of $O_s$ is $\left[M * \left(M+1\right) / 2\right]*d_M$.

\paragraph{\textbf{Cross feature combination}}
We use an extra cross-layer to extract interactions between different structures of user information:
\begin{equation}
    O^u_2=\text{Cross}^u_1\left(O^u_1, O^u_1\right)
\end{equation}
where $O_1^u=[O^d, O^s]$ is the concatenation of dense and sparse features. Finally, we concatenate $O_1^u$ and $O_2^u$ as $O^u$ and then use a linear layer to combine them as the final cross feature:
\begin{equation}
    f^u_c = ({O^u})^T \cdot W^u + b^u
\end{equation}

\subsubsection{\textbf{Integrate MLP}}\par
Besides feature crossing, we also adopt a two-layer MLP to implicitly capture interactions between different feature fields, and then we integrate cross features and MLP features as the final user representation:
\begin{equation}\label{eq:encu}
   \begin{aligned}
    f^u_d &= \text{MLP}(\left[D,E\right]) \\
    f^u &= \left[f^u_c, f^u_d\right]
\end{aligned} 
\end{equation}

where $f^u \in \mathcal{R}^{d_N}$ is in the common hidden space of visual and text features.

\subsection{Popularity Predictor}
After extracting post and user features,  we combine them to learn the popularity representation by using the two-layer MLP as a popularity encoder, and a linear layer is adopted to make final regression:
\begin{align}
    f^{pop} &= \text{MLP}(\left[f^p, f^u\right]) \label{eq:encpop}\\
    y &= {f^{pop}}^T \cdot W + b
\end{align}
where $f^{pop} \in \mathcal{R}^{d_h}, W \in \mathcal{R}^{d_h \times 1},b \in \mathcal{R}$. We use the Mean Square Error (MSE) as the loss function to supervise the regression process:
\begin{equation}
   \mathcal{L}_{reg} = \sum_{i=1}^{n}{\left(\hat{y_i}-y_i\right)}^2 
\end{equation}
where $n$ is the number of samples, $y_i$ is the predicted value of sample $i$ and $\hat{y_i}$ is the corresponding ground-truth. With the supervised signal of popularity label, $f^{pop}$ as the output of the last hidden layer of the network can learn features most associated with popularity.

\section{Contrastive Learning Tasks}
So far, we have introduced the main architecture of our model. However, it still only encodes post and user information given by the dataset, ignoring implicit social factors that affect the popularity of posts. We have identified three potential social factors, i.e., content relevance, user influence similarity, and user identity, thereby we design three supervised contrastive learning tasks to inject these factors into the model. We begin with a brief description of the Supervised Contrastive Learning (SupCon) loss ~\cite{khosla2021supervised}.

\subsection{Supervised Contrastive Learning}
Given $N$ paired samples and their corresponding labels in a mini-batch $\mathcal{D}=\{{\hat{x}_i};\hat{y_i}\}_{i=1}^{2N}$, $\hat{x}_{2k}$ and $\hat{x}_{2k+1}$ represent a positive pair obtained through augmentation, thus sharing the same label, i.e., $\hat{y}_{2k}=\hat{y}_{2k+1}$. We denote $I=\{1,...,2N\}$ as the index of all samples, and $j(i)$ be the index of a positive sample associated with $\hat{x}_i$, (e.g., if $i=2k$, then $j(i)=2k+1$), $A(i) = I \backslash \{i\}$ is all index except $i$, then $P(i)=\{p \in A(i) | \hat{y_p} = \hat{y_i}\}$ indicate the samples which have the same label with $\hat{x}_i$, i.e., all positive samples in a mini-batch, the other samples in the same batch are treated as negative samples. We can calculate the SupCon objective for all data in a mini-batch as follows:
\begin{equation} \label{eq:supcon}
    \mathcal{L}^{sup} = -\sum_{i \in I} \frac{1}{|P(i)|} \sum_{p \in P(i)} log \frac{e^{\operatorname{sim}\left(\mathbf{f}_i, \mathbf{f}_p\right) / \tau}}{\sum_{a \in A(i)} e^{\operatorname{sim}\left(\mathbf{f}_i, \mathbf{f}_a\right) / \tau}} 
\end{equation}
where $f_i$ is the feature of $\hat{x}_i$, $\tau$ is the temperature hyperparameter and $\operatorname{sim}(f_1, f_2)$ is the cosine similarity $\frac{f_1^T\cdot f_2}{\Vert f_1\Vert \cdot \Vert f_2\Vert}$. This supervised loss encourages the encoder to give closely aligned representations to all samples from the same class.\par

\subsection{Content Relevance Discrimination Task} 
Influenced by user preferences on social platforms, posts with correlated content are likely to have more similar audiences. Therefore, we propose the Content Relevance Discrimination (CRD) task for the Post Encoder ($Enc_p$) to pull representations of posts with correlated content closer while pushing posts with different content away. Therefore, we use category labels as the supervised signal to represent correlations of post content. The category labels can be hierarchical, e.g., animal-dog-hound to characterize the differences in post content at a granular level. We denote the number of hierarchical levels as $L$, and $l \in L$ is a level in the multi-label, then we adopt Hierarchial Multi-label Contrastive Learning loss ~\cite{zhang2022use}, which extends Eq. \ref{eq:supcon} to the multi-label scenario, as the objective function. Similar to $P(i)$, we denote $P_l(i) = \{p \in A(i) | \hat{y}^l_p = \hat{y}^l_i\}$ as the samples which have the same category label at level $l$ with $\hat{x}_i$. Then the training objective of the CRD task is:
\begin{equation}
    \mathcal{L}_{CRD} = -\sum_{l \in L} \frac{1}{|L|} \sum_{i \in I} \frac{\lambda_l}{|P_l(i)|} \sum_{p_l \in P_l} log \frac{e^{\operatorname{sim}\left(\mathbf{f}_i^p, \mathbf{f}_{p_l}^p\right) / \tau}}{\sum_{a \in A(i)} e^{\operatorname{sim}\left(\mathbf{f}_i^p, \mathbf{f}_a^p\right) / \tau}}
\end{equation}
The parameter $\lambda_l = F(l)$ functions as a penalty parameter, imposing a higher penalty on sample pairs at higher levels. This is because positive samples at lower levels may give rise to negative sample pairs at higher levels. Here, we use $\lambda_l = 1/l^2$ as the penalty function. Since the hierarchical labels are used to describe the post, i.e. image and text, we use $f^p_i$, the final output through $Enc_p$ computed by Eq. \ref{eq:pmlp}, as the feature of the sample $\hat{x}_i$ to compute $\mathcal{L}_{CRD}$. Thus the gradient can be computed as $\nabla \mathcal{L}_{CRD}=\frac{\partial \operatorname{\mathcal{L}_{CRD}}}{\partial \Theta_{Enc_p}}$, which will flow through all the parameters of the Post Encoder, allowing it to discriminate between different categories of post content. 

\subsection{User Influence Similarity Discrimination Task}
On social platforms, more influential users attract more attention. Therefore, we design the User Influence Similarity Discrimination (UISD) task to encourage the user encoder ($Enc_u$) to pull representations of users with similar influence closer, and vice versa to push them away. We use the $FollowerCount$ attribute to represent the influence size of a user, since intuitively users with more followers will also have more influence. However, $FollowerCount \in \mathcal{R}^+$ is difficult to use directly as a target feature for contrastive learning. To this end, we utilize a clustering algorithm to assign a label of discrete influence level to each user. Specifically, we adopt K-means to cluster all data samples into $k$ clusters according to their $FollowerCount$ attribute, then for the sample $\hat{x}_i$ in the cluster $c_j$, its label of user influence level is assigned as $\hat{y}_i^c=j$. Then we can denote positive sample set of $\hat{x}_i$ as $P_c(i)=\{p \in A(i) | \hat{y}^c_p = \hat{y}^c_i\}$, and adopt Eq. \ref{eq:supcon} as the training objective of the UISD task:
\begin{equation}
    \mathcal{L}_{UISD} = -\sum_{i \in I} \frac{1}{|P_c(i)|} \sum_{p \in P_c(i)} log \frac{e^{\operatorname{sim}\left(\mathbf{f}_i^u, \mathbf{f}_p^u\right) / \tau}}{\sum_{a \in A(i)} e^{\operatorname{sim}\left(\mathbf{f}_i^u, \mathbf{f}_a^u\right) / \tau}} 
\end{equation}
where $f_i^u$ is the final output of the sample $\hat{x}_i$ through $Enc_u$ computed by Eq. \ref{eq:encu}. The gradient can be computed as $\nabla \mathcal{L}_{UISD}=\frac{\partial \operatorname{\mathcal{L}_{UISD}}}{\partial \Theta_{Enc_u}}$, which will flow through all the parameters of the User Encoder, allowing it to discriminate between posts published by users with different influence levels.

\subsection{User Identity Discrimination Task}
As a user's followers tend to consistently engage with nearly every post from that user, the popularity of a post created by that user should exhibit a higher correlation with the popularity of their other posts than with the posts made by different users. Therefore, we propose the User Identity Discrimination (UID) task to encourage the popularity encoder $(Enc_{pop})$ to discriminate the final popularity feature of posts from different users. For sample $\hat{x}_i$, we use the $UserId$ attributes which identify each individual user as the label $\hat{y}_i^u$, then we denote positive sample set $P_u(i)=\{p \in A(i) | \hat{y}^u_p = \hat{y}^u_i\}$ and utilize Eq. \ref{eq:supcon} as the training objective of the UID task:
\begin{equation}
    \mathcal{L}_{UID} = -\sum_{i \in I} \frac{1}{|P_u(i)|} \sum_{p \in P_u(i)} log \frac{e^{\operatorname{sim}\left(\mathbf{f}_i^{pop}, \mathbf{f}_p^{pop}\right) / \tau}}{\sum_{a \in A(i)} e^{\operatorname{sim}\left(\mathbf{f}_i^{pop}, \mathbf{f}_a^{pop}\right) / \tau}} 
\end{equation}
where $f_i^{pop}$ is the final output of $\hat{x}_i$ through $Enc_{pop}$ computed by Eq. \ref{eq:encpop}. The gradient can be computed as $\nabla \mathcal{L}_{UID}=\frac{\partial \operatorname{\mathcal{L}_{UID}}}{\partial \Theta_{Enc_{pop}}}$, which only flows through parameters of the MLP in Eq. \ref{eq:encpop} so that the UID task focuses on tuning the feature space of the final popularity features without interfering with the optimization process of the previous Post and User encoders.

\subsection{Unsupervised Augmentation}
Remember that for each sample $\hat{x}_i$, there is a positive sample $\hat{x}_{j(i)}$ which is the augmentation of $\hat{x}_i$. The augmentation methods include masking, cropping, and reordering for images ~\cite{dosovitskiy2015discriminative}, and a similar approach is used for natural language ~\cite{wu2020clear}. However, in our model, the three contrastive learning tasks optimize three encoders using inputs from diverse data sources, including images, text, numerical, categorical data, and their combinations. Therefore, we want to use a unified augmentation approach to provide high semantic similarity for positive pairs under various data sources. Inspired by ~\cite{gao2022simcse, Qiu_2022}, we adopt an unsupervised model-level augmentation method by Dropout masks in the model. Notice that all three encoders optimized by our contrastive learning tasks contain MLPs, so we can add Dropout modules after the linear layer of the MLP. By feeding the features twice with different Dropout masks in the forward-passing process, semantically similar but different outputs can be obtained:
\begin{align}
    f_i^t &= \text{Dropout}(Enc_t(z^t_i);\theta_1) \\
    f_{j(i)}^t &= \text{Dropout}(Enc_t(z^t_i);\theta_2)
\end{align}
where $t$ denotes different types of encoders, i.e., $Enc_p$, $Enc_u$, and $Enc_{pop}$, $z^t_i$ is the input value of $Enc_t$ corresponding to sample $\hat{x}_i$, $\theta_1$ and $\theta_2$ denote different Dropout masks of two feed-forward processes.

\subsection{Batch Sampling Strategy}
The importance of sampling for representation learning has been shown in ~\cite{khosla2021supervised,wu2018sampling}. Due to the utilization of different labels in our three supervised contrastive learning tasks, it is not feasible to ensure that there are enough positive and negative samples for all tasks in a batch of data. Therefore, our sampling method is designed for the CRD task, as optimizing the hierarchical loss imposes the strictest requirements on sample pairs, demanding that each sample can form a positive pair with samples that share a common ancestry at all levels in the structure. Denote the batch size is $B$ and the hierarchical levels are $\{1,...,l\}$, then our method is to sample $(l + 1)$ blocks $\{b_0,b_1,...,b_l\}$ that $|b_i|= B/(l+1)$. First, we randomly sample $b_{0}$ to ensure sample diversity for each label. Then we repeat this operation: randomly sample $b_i$ such that the label of each sample in $b_i$ is the same as the label of the corresponding sample in $b_{0}$ on $\text{level}_{i}$, but different on $\text{level}_{i+1}$, where $i=\{1,2,...,l\}$. After $l$ iterations, we get a batch $\{b_0,b_1,...,b_l\}$ that ensures sufficient representation from all levels of the hierarchy for each anchor sample.

\begin{table*}[t]
    \centering
    \begin{tabular}{lccccccccc}
        \toprule \multirow{2}{*}{ Method } & \multicolumn{3}{c}{ SMPD-$100K$ } & \multicolumn{3}{c}{ SMPD-$200K$ } & \multicolumn{3}{c}{ SMPD-$300K$ } \\
        \cline { 2 - 10 } & MAE & MSE & SRC & MAE & MSE & SRC & MAE & MSE & SRC \\
        \midrule
        DTCN ~\cite{wu2017sequential} & $1.606$ & $4.011$ & $0.602$ & $1.579$ & $3.997$ & $0.613$ & $1.532$ & $3.573$ & $0.624$ \\
        Att-MLP ~\cite{xu2020multimodal} & $1.576$ & $3.642$ & $0.617$ & $1.535$ & $3.568$ & $0.621$ & $1.453$ & $3.432$ & $0.635$ \\
        Multiview ~\cite{tan2022efficient} & $1.417$ & $3.383$ & $0.672$ & $1.409$ & $3.411$ & $0.679$ & $1.387$ & $3.325$ & $0.693$ \\
        TTC-VLT ~\cite{chen2022and} & $1.413$ & $3.378$ & $0.683$ & $1.401$ & $3.354$ & $0.687$ & $1.346$ & $3.274$ & $0.711$ \\
        DSN ~\cite{zhang2023improving} & $1.394$ & $3.336$ & $0.661$ & $1.389$ & $3.270$ & $0.683$ & $1.292$ & $2.954$ & $0.714$ \\
        VisualR ~\cite{LIU2023103738} & $1.343$ & $3.310$ & $0.689$ & $1.291$ & $2.920$ & $0.716$ & $1.207$ & $2.605$ & $0.733$ \\
        DFT-MOVLT ~\cite{chen2023} & $1.295$ & $3.007$ & $0.701$ & $1.278$ & $2.883$ & $0.718$ & $1.198$ & $2.528$ & $0.739$ \\
        \midrule
        RNC ~\cite{zha2023rankncontrast} & $1.251$ & $2.785$ & $0.724$ & $1.216$ & $2.658$ & $0.737$ & $1.205$ & $2.532$ & $0.749$ \\
        \midrule
        \textbf{PPCL (Ours)} & $\mathbf{1.197}^*$ & $\mathbf{2.499}^*$ & $\mathbf{0.737}$ & $\mathbf{1.195}^*$ & $\mathbf{2.522}^*$ & $\mathbf{0.741}$  & $\mathbf{1.164}^*$ & $\mathbf{2.374}^*$ & $\mathbf{0.762}$\\
        \bottomrule
    \end{tabular}
    \caption{Overall comparison between baselines and PPCL on three datasets, measured by MAE, MSE, and SRC. The best results are in \textbf{bold}. A paired t-test is performed on MAE and MSE, and $*$ indicates a statistical significance $p < 0.05$ as compared to the best baseline.}
    \label{tab:comparison}
    \vspace{-0.5cm}
\end{table*}

\subsection{Joint Optimization}
To jointly optimize the main popularity prediction task and three contrastive learning auxiliary tasks, we design a weighted loss function:
\begin{equation}
    \mathcal{L} = \lambda \mathcal{L}_{reg} + (1-\lambda)(\alpha_1 \mathcal{L}_{CRD} + \alpha_2 \mathcal{L}_{UISD} + \alpha_3 \mathcal{L}_{UID})
\end{equation}
where $\lambda$ is the hyperparameter to control the rate of main and auxiliary tasks. Since three contrastive learning tasks aim to optimize different encoders with different data sources, we use $\alpha_1,\alpha_2,\alpha_3$ to regulate their convergence rate.

\section{experiment}


\subsection{Setup}

\subsubsection{\textbf{Datasets}}
We use the Social Media Popularity Dataset\footnote{\url{https://smp-challenge.com/2023/index.html}} ~\cite{wu2019smp} (SMPD) collected from Flickr, which is widely used by previous works, to evaluate the performance of our method. SMPD includes $305,595$ posts published by $38,307$ users. The posts were published between 2015-03 and 2016-03, and the viewing numbers of these posts were recorded in 2016-07. The hierarchical category information in SMPD is displayed in $3$ levels, i.e., Category, Subcategory, and Concept, with the number of labels at each level being $11$, $77$, and $668$, respectively. Following previous work, we also use user information provided by Hyfea ~\cite{lai2020hyfea}, which complements the data from SMPD. Since the amount of data available varies in different application scenarios, we randomly sample the original SMPD dataset and obtain three datasets of sizes $100K$, $200K$, and $300K$, to test the performance of models with different data volumes. Details of datasets are shown in Appendix ~\ref{app:user}. Each dataset is naturally sorted by posting time and divided into training, validation, and test sets in a ratio of $8$:$1$:$1$. The experimental results are reported on the test set.\par

\subsubsection{\textbf{Metrics}}
To evaluate the prediction performance, we use two precision metrics, Mean Absolute Error (MAE) and Mean Square Error (MSE), and a correlation metric Spearman Ranking Correlation (SRC) as in ~\cite{wu2017sequential,LIU2023103738}. Lower MAE and MSE / higher SRC refer to better performance.

\subsubsection{\textbf{Baselines}}
We use the following social media popularity prediction methods for comparisons:
\begin{itemize}
    \item \textbf{DTCN} ~\cite{wu2017sequential} uses ResNet to extract image features, and adopts a time-aware attention method to aggregate publishing time-related posts.
    \item \textbf{Att-MLP} ~\cite{xu2020multimodal} combines ResNet and Word2Vec for image and text features, adopting feature-level attention to dynamically aggregate different features.
    \item \textbf{Multiview} ~\cite{tan2022efficient} utilizes ALBEF to encode image and text, employing time transformers to aggregate users' historical posts.
    \item \textbf{TTC-VLT} ~\cite{chen2022and} uses ViLT and designs a title-tag contrastive learning task for better text representation.
    \item \textbf{DSN} ~\cite{zhang2023improving} utilizes CLIP and designs content-aware attention to aggregate content-related posts.
    \item \textbf{VisualR} ~\cite{LIU2023103738} combines Faster R-CNN and a pre-trained scene graph generator model to extract relationships between objects in the image.
    \item \textbf{DFT-MOVLT} ~\cite{chen2023} conducts three multimodal pre-training tasks to enhance ALBEF and designs an extra classification task to improve the results of the regression.
\end{itemize}
In addition, we select a contrastive regression method, \textbf{Rank-N-Contrast (RNC)} ~\cite{zha2023rankncontrast} for comparison. RNC is a framework designed for regression tasks that learns continuous representations by contrasting samples against each other based on their rankings in the target space. Because the implementation of RNC is a contrastive loss orthogonal to any regression model, we incorporate it into our base architecture, replacing our three contrastive tasks, to compare the contrastive learning schemes we have devised. We provide details of all baseline implementations in Appendix ~\ref{app:baseline}.

\subsubsection{\textbf{Implementation}}
For all three datasets, the training batch size is $512$ and the learning rate is $0.002$. Adam is the training optimizer and weight decay of $0.0001$ is used to avoid over-fitting. The number of categorical features and numerical features are $9$ and $11$, the details of user features are shown in Appendix ~\ref{app:user}. The embedding size is $32$ and the hidden size is $512$. The number of dense feature cross-layers is $4$. The activation function used in MLPs is ReLU and the dropout rate is $0.1$. The temperature $\tau$ is $0.1$, and the number of clusters $k$ in the UISD task is $40$. The loss weights $\lambda,\alpha_1,\alpha_2,\alpha_3$ are set to $0.9, 3, 1, 0.1$. Since our design freezes the parameters of CLIP, we pre-extract $f_r^v$ and $f_r^t$, and feed them directly into the model during the training process. To achieve the best performance, we train until the loss stops dropping for 10 consecutive epochs. Our method is implemented with PyTorch and the experiment is conducted on one NVIDIA 3090 GPU.

\subsection{Overall Comparisons}
We compare our method PPCL with eight strong baselines, and the results on three datasets are shown in Table ~\ref{tab:comparison}. Our PPCL achieves the best results under all three metrics. Compared to the optimal popularity prediction baseline, PPCL obtains MAE improvements of $7.6\%$, $6.5\%$, and $2.8\%$ on SMPD-$100K$, $200K$, $300K$, respectively. \par
Among all baselines, DTCN and Att-MLP perform the worst, which is due to their insufficient feature extraction for text and images. Compared to them, Multiview, TTC-VLT and DSN all place more emphasis on modeling the multimodal content of posts by employing pre-trained models and different fine-tuning methods, and therefore achieve better results. VisualR goes a step further in understanding the content of the post and achieves better results by understanding the relationship of objects in the image. DFT-MOVLT is the only method that adopts multimodal pre-training, leading to the best performance of all baselines. The direction of improvement in the popularity prediction baselines centers around how to enhance the presentation of image and text features. The content of a post is undoubtedly important, but the excessive reliance on the development of pre-trained models for content extraction is notable. In our approach, we only use a simple adapter to fine-tune the features extracted by CLIP. For the first time, we propose to consider the impact of underlying social factors on popularity and subtly design contrastive learning tasks to model these factors. This approach has allowed us to achieve competitive results. Compared to the contrastive learning baseline RNC, as it is a general contrastive loss designed for regression tasks and lacks consideration for social factors in popularity prediction, its performance is also inferior to our specific-designed contrastive learning scheme.

Another valuable finding is the data efficiency; that is, the smaller the dataset, the greater the improvement in the performance of PPCL. We infer that this is attributed to the proposed social factors, which serve as inductive biases because they encapsulate prior knowledge about the drivers of popularity in social media contexts. Incorporating these social factors allows PPCL to achieve better results without requiring a large amount of training data.

\subsection{Contrastive Learning Analysis}
In this section, we focus on analyzing the performance of the contrastive learning module in PPCL. More experiments about parameter sensitivity and case study are provided in Appendix \ref{app:param} \& \ref{app:case}.
\subsubsection{Effectiveness of contrastive tasks}
To verify the effectiveness of three contrastive learning tasks, we design the ablation experiment as shown in Table ~\ref{tab:ablation}.
We start with model w/o CL, i.e., no contrastive learning task is added, the model's effectiveness on all three datasets decreases significantly, i.e., compared to PPCL, the MAE increases by $6.9\%$, $3.8\%$, and $3.6\%$, respectively. This indicates that the performance of the model can be effectively improved in the case of joint learning of the three contrastive tasks.

Using w/o CL as a baseline, we add different contrastive tasks on top to analyze the effectiveness of each task on different datasets. For SMPD-$300K$ and $200K$, the $\mathcal{L}_{UID}$ task provides the most significant model enhancement. This is due to the strong correlation between user identity and popularity, which also suggests that the UID task makes the model learn this pattern well. However, when the data set size is reduced to $100K$, the effectiveness of $\mathcal{L}_{UID}$ decreases most severely, while $\mathcal{L}_{UISD}$ becomes the best. We begin by noting that for all three contrastive tasks, a decrease in dataset size will lead to a reduction in the number of positive samples. This may have varying degrees of impact on the effectiveness of these tasks. For the $\mathcal{L}_{UID}$ task, since $Enc_{pop}$ is the closest to the model's output layer, fluctuations in its effect can also have a large impact on the prediction results. For the $\mathcal{L}_{CRD}$ task, since the supervised label it uses is hierarchical, it is also more affected by the reduction of positive samples. As a result, the $\mathcal{L}_{UISD}$ task uses simpler, single-level labels while at the front of the model, so that its decrease in effectiveness is relatively less detrimental to performance. 

\subsubsection{Effectiveness of sampling and augmentation}
We also test the effectiveness of our sampling and augmentation methods. For the sampling method, we use random sampling instead. This will result in an unstable number of positive samples, and in the worst case, there may be no positive samples. This makes it difficult for the model to correctly learn the semantic associations of similar samples, leading to worse performance. For augmentation, we stop using dropout to generate positive sample $\hat{x}_{j(i)}$. This can lead to a large reduction in the number of positive/negative samples, making the model less discriminating and generating less expressive features, eventually leading to performance degradation.

\begin{table}[t]
    \centering
    \begin{tabular}{lcccccc}
    \toprule \multirow{2}{*}{ Method } & \multicolumn{2}{c}{ SMPD-$100K$ } & \multicolumn{2}{c}{ SMPD-$200K$ } & \multicolumn{2}{c}{ SMPD-$300K$ } \\
        \cline { 2 - 7 } & MAE & SRC & MAE & SRC & MAE & SRC \\
        \midrule
        w/o CL & $1.280$ & $0.708$ & $1.241$ & $0.731$ & $1.206$ & $0.738$ \\
        + $\mathcal{L}_{CRD}$ & $1.254$ & $0.722$ & $1.219$ & $0.733$ & $1.187$ & $0.741$ \\
        + $\mathcal{L}_{UISD}$ & $1.230$ & $0.733$ & $1.217$ & $0.736$ & $1.190$ & $0.741$ \\
        + $\mathcal{L}_{UID}$ & $1.274$ & $0.713$ & $1.214$ & $0.737$ & $1.177$ & $0.749$  \\
        + $\mathcal{L}_{CRD} + \mathcal{L}_{UISD}$  & $1.235$ & $0.728$ & $1.215$ & $0.730$ & $1.183$ & $0.745$ \\
        + $\mathcal{L}_{CRD} + \mathcal{L}_{UID}$  & $1.264$ & $0.718$ & $1.212$ & $0.741$ & $1.182$ & $0.747$ \\
        + $\mathcal{L}_{UISD} + \mathcal{L}_{UID}$  & $1.229$ & $0.724$ & $1.204$ & $0.741$ & $1.174$ & $0.755$ \\
        \midrule
        Random Sampling & $1.230$ & $0.722$ & $1.212$ & $0.734$ & $1.189$ & $0.743$\\
        w/o Augmentation & $1.259$ & $0.720$ & $1.210$ & $0.735$ & $1.198$ & $0.747$ \\ 
        \midrule 
        PPCL & 1.197 & 0.737 & 1.195 & 0.741 & 1.164 & 0.762 \\
    \bottomrule
    \end{tabular}
    \caption{Ablation analysis of contrastive learning modules.}
    \label{tab:ablation}
    \vspace{-0.8cm}
\end{table}

\subsubsection{Impact of negative samples}
To test the impact of negative samples, we provide a comparison between using different training batch sizes. As the batch size increases, the number of negative samples within the batch also increases. Table \ref{tab:batch} shows that the predictions of our model are relatively stable when batch size $\geq 64$. However, when the batch size is particularly small (i.e., $32$ or $16$), the performance of the model decreases dramatically, meaning that too few negative samples could impair the representation capacity. 

\begin{table}[tb]
    \centering
    \begin{tabular}{lcccccc}
\toprule \multirow{2}{*}{ Batch size } & \multicolumn{2}{c}{ SMPD-$100K$ } & \multicolumn{2}{c}{ SMPD-$200K$ } & \multicolumn{2}{c}{ SMPD-$300K$ } \\
        \cline { 2 - 7 } & MAE & SRC & MAE & SRC & MAE & SRC \\
        \midrule
16 &1.264&0.712	&1.335&0.681&1.344&0.683 \\
32	&1.258&0.718&	1.206&0.742&	1.299&	0.697\\
64	&1.202&$\mathbf{0.740}$&	1.201&$\mathbf{0.743}$	&1.191&0.744\\
128&	1.224&0.728&	1.217&0.742&	1.218&0.735\\
256&	1.227&0.725	&1.216&0.741&	1.181&0.753\\ 
512	& $\mathbf{1.197}$&0.737&	$\mathbf{1.195}$	&0.741	&$\mathbf{1.164}$&	$\mathbf{0.762}$\\
\bottomrule
    \end{tabular}
    \caption{Sensitivity to different batch sizes.}
    \label{tab:batch}
    \vspace{-0.5cm}
\end{table}

\begin{figure}[tb]
    \centering
    \includegraphics[width=\linewidth]{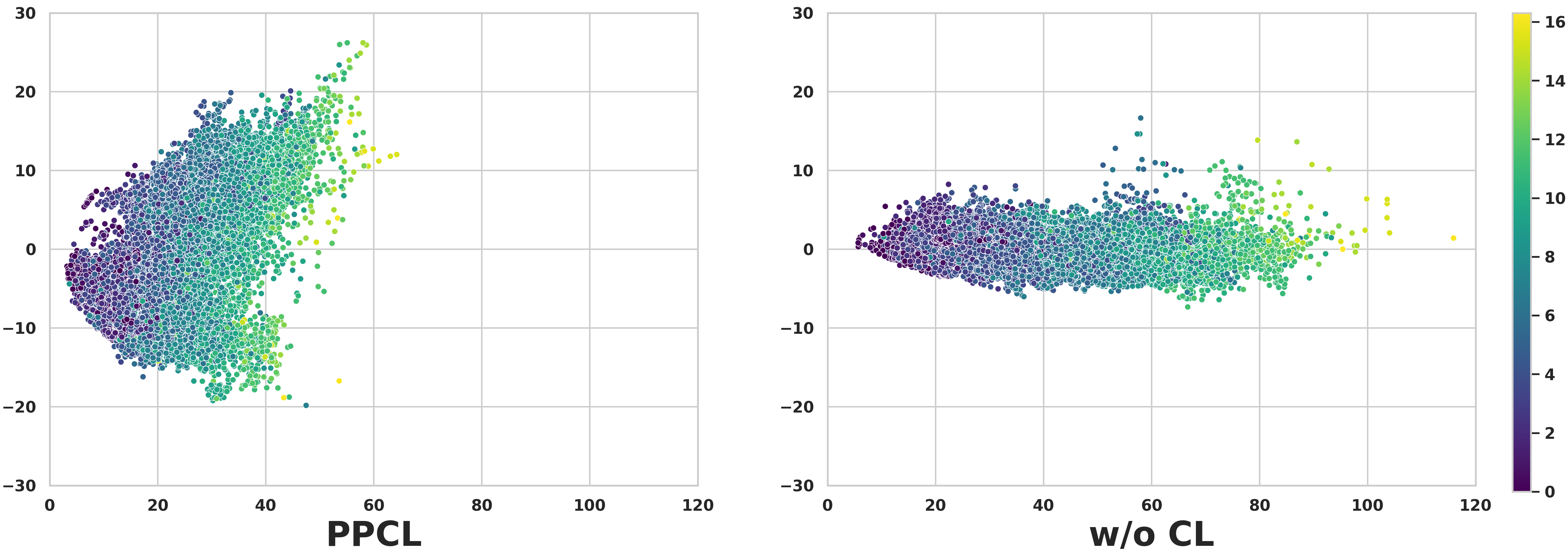}
    \caption{Visualization of features output by $Enc_{pop}$ of PPCL and w/o CL with colors indicating values of post popularity.}
    \label{fig:vis}
    \vspace{-0.5cm}
\end{figure}

\subsection{Feature Visualization}\label{app:visual}


To validate the representation capacity of our contrastive scheme, we use SVD to project popularity features $f^{pop}$ learned by PPCL and w/o CL on the test set of SMPD-$300K$ into a 2-dimensional space. As shown in Figure ~\ref{fig:vis}, different colors indicate the value of popularity labels. The features generated by w/o CL exhibit a narrow distribution along the y-axis, indicating that the model faces the representation degeneration problem ~\cite{ethayarajh-2019-contextual,gao2019representation}, making the features of posts with different popularity less discriminative. On the contrary, features generated by PPCL, owing to the contrastive learning tasks, demonstrate a more uniform distribution. The uniformity in feature space is highly related to the model performance, as shown in ~\cite{wang2022contra}, which suggests PPCL has a stronger representation capability, eventually resulting in improved prediction performance.
\section{Conclusion}
In this paper, we investigate the important effect of implicit social factors on post popularity that has not been studied in previous social media popularity prediction work. By analyzing users' post-browsing habits, we propose three implicit social factors, the analysis of the dataset also validates our hypothesis. Based on this, we design three contrastive learning tasks and corresponding sampling and augmentation algorithms, which enable the model to jointly optimize the contrastive losses for different encoders as well as different data sources. On the Social Media Popularity Prediction Dataset, comprehensive experiments conducted under three different settings have demonstrated PPCL's superior predictive performance, data efficiency, and strong representation capabilities. In the future, we will further optimize the contrastive framework, including supervised label selection, sample augmentation and sampling process, to better model the impact of social factors on popularity.
\bibliographystyle{ACM-Reference-Format}
\bibliography{sample-base}


\begin{thebibliography}{43}


\ifx \showCODEN    \undefined \def \showCODEN     #1{\unskip}     \fi
\ifx \showDOI      \undefined \def \showDOI       #1{#1}\fi
\ifx \showISBNx    \undefined \def \showISBNx     #1{\unskip}     \fi
\ifx \showISBNxiii \undefined \def \showISBNxiii  #1{\unskip}     \fi
\ifx \showISSN     \undefined \def \showISSN      #1{\unskip}     \fi
\ifx \showLCCN     \undefined \def \showLCCN      #1{\unskip}     \fi
\ifx \shownote     \undefined \def \shownote      #1{#1}          \fi
\ifx \showarticletitle \undefined \def \showarticletitle #1{#1}   \fi
\ifx \showURL      \undefined \def \showURL       {\relax}        \fi
\providecommand\bibfield[2]{#2}
\providecommand\bibinfo[2]{#2}
\providecommand\natexlab[1]{#1}
\providecommand\showeprint[2][]{arXiv:#2}

\bibitem[Cao et~al\mbox{.}(2017)]%
        {cao2017deephawkes}
\bibfield{author}{\bibinfo{person}{Qi Cao}, \bibinfo{person}{Huawei Shen}, \bibinfo{person}{Keting Cen}, \bibinfo{person}{Wentao Ouyang}, {and} \bibinfo{person}{Xueqi Cheng}.} \bibinfo{year}{2017}\natexlab{}.
\newblock \showarticletitle{DeepHawkes: Bridging the Gap between Prediction and Understanding of Information Cascades}. In \bibinfo{booktitle}{\emph{Proceedings of the 2017 ACM on Conference on Information and Knowledge Management}}. \bibinfo{pages}{1149–1158}.
\newblock


\bibitem[Chen et~al\mbox{.}(2019)]%
        {chen2019social}
\bibfield{author}{\bibinfo{person}{Junhong Chen}, \bibinfo{person}{Dayong Liang}, \bibinfo{person}{Zhanmo Zhu}, \bibinfo{person}{Xiaojing Zhou}, \bibinfo{person}{Zihan Ye}, {and} \bibinfo{person}{Xiuyun Mo}.} \bibinfo{year}{2019}\natexlab{}.
\newblock \showarticletitle{Social Media Popularity Prediction Based on Visual-Textual Features with XGBoost}. In \bibinfo{booktitle}{\emph{Proceedings of the 27th {ACM} International Conference on Multimedia, {MM} 2019, Nice, France, October 21-25, 2019}}. \bibinfo{publisher}{{ACM}}, \bibinfo{pages}{2692--2696}.
\newblock
\urldef\tempurl%
\url{https://doi.org/10.1145/3343031.3356072}
\showDOI{\tempurl}


\bibitem[Chen et~al\mbox{.}(2022)]%
        {chen2022and}
\bibfield{author}{\bibinfo{person}{Weilong Chen}, \bibinfo{person}{Chenghao Huang}, \bibinfo{person}{Weimin Yuan}, \bibinfo{person}{Xiaolu Chen}, \bibinfo{person}{Wenhao Hu}, \bibinfo{person}{Xinran Zhang}, {and} \bibinfo{person}{Yanru Zhang}.} \bibinfo{year}{2022}\natexlab{}.
\newblock \showarticletitle{Title-and-Tag Contrastive Vision-and-Language Transformer for Social Media Popularity Prediction}. In \bibinfo{booktitle}{\emph{{MM} '22: The 30th {ACM} International Conference on Multimedia, Lisboa, Portugal, October 10 - 14, 2022}}. \bibinfo{publisher}{{ACM}}, \bibinfo{pages}{7008--7012}.
\newblock
\urldef\tempurl%
\url{https://doi.org/10.1145/3503161.3551568}
\showDOI{\tempurl}


\bibitem[Chen et~al\mbox{.}(2023)]%
        {chen2023}
\bibfield{author}{\bibinfo{person}{Xiaolu Chen}, \bibinfo{person}{Weilong Chen}, \bibinfo{person}{Chenghao Huang}, \bibinfo{person}{Zhongjian Zhang}, \bibinfo{person}{Lixin Duan}, {and} \bibinfo{person}{Yanru Zhang}.} \bibinfo{year}{2023}\natexlab{}.
\newblock \showarticletitle{Double-Fine-Tuning Multi-Objective Vision-and-Language Transformer for Social Media Popularity Prediction}. In \bibinfo{booktitle}{\emph{Proceedings of the 31st {ACM} International Conference on Multimedia, {MM} 2023, Ottawa, ON, Canada, 29 October 2023- 3 November 2023}}. \bibinfo{publisher}{{ACM}}, \bibinfo{pages}{9462--9466}.
\newblock
\urldef\tempurl%
\url{https://doi.org/10.1145/3581783.3612845}
\showDOI{\tempurl}


\bibitem[Ding et~al\mbox{.}(2019)]%
        {ding2019social}
\bibfield{author}{\bibinfo{person}{Keyan Ding}, \bibinfo{person}{Ronggang Wang}, {and} \bibinfo{person}{Shiqi Wang}.} \bibinfo{year}{2019}\natexlab{}.
\newblock \showarticletitle{Social Media Popularity Prediction: {A} Multiple Feature Fusion Approach with Deep Neural Networks}. In \bibinfo{booktitle}{\emph{Proceedings of the 27th {ACM} International Conference on Multimedia, {MM} 2019, Nice, France, October 21-25, 2019}}. \bibinfo{publisher}{{ACM}}, \bibinfo{pages}{2682--2686}.
\newblock
\urldef\tempurl%
\url{https://doi.org/10.1145/3343031.3356062}
\showDOI{\tempurl}


\bibitem[Dosovitskiy et~al\mbox{.}(2016)]%
        {dosovitskiy2015discriminative}
\bibfield{author}{\bibinfo{person}{Alexey Dosovitskiy}, \bibinfo{person}{Philipp Fischer}, \bibinfo{person}{Jost~Tobias Springenberg}, \bibinfo{person}{Martin~A. Riedmiller}, {and} \bibinfo{person}{Thomas Brox}.} \bibinfo{year}{2016}\natexlab{}.
\newblock \showarticletitle{Discriminative Unsupervised Feature Learning with Exemplar Convolutional Neural Networks}.
\newblock \bibinfo{journal}{\emph{{IEEE} Trans. Pattern Anal. Mach. Intell.}} \bibinfo{volume}{38}, \bibinfo{number}{9} (\bibinfo{year}{2016}), \bibinfo{pages}{1734--1747}.
\newblock
\urldef\tempurl%
\url{https://doi.org/10.1109/TPAMI.2015.2496141}
\showDOI{\tempurl}


\bibitem[Ethayarajh(2019)]%
        {ethayarajh-2019-contextual}
\bibfield{author}{\bibinfo{person}{Kawin Ethayarajh}.} \bibinfo{year}{2019}\natexlab{}.
\newblock \showarticletitle{How Contextual are Contextualized Word Representations? Comparing the Geometry of BERT, ELMo, and {GPT-2} Embeddings}. In \bibinfo{booktitle}{\emph{Proceedings of the 2019 Conference on Empirical Methods in Natural Language Processing and the 9th International Joint Conference on Natural Language Processing, {EMNLP-IJCNLP} 2019, Hong Kong, China, November 3-7, 2019}}. \bibinfo{publisher}{Association for Computational Linguistics}, \bibinfo{pages}{55--65}.
\newblock
\urldef\tempurl%
\url{https://doi.org/10.18653/V1/D19-1006}
\showDOI{\tempurl}


\bibitem[Gao et~al\mbox{.}(2019)]%
        {gao2019representation}
\bibfield{author}{\bibinfo{person}{Jun Gao}, \bibinfo{person}{Di He}, \bibinfo{person}{Xu Tan}, \bibinfo{person}{Tao Qin}, \bibinfo{person}{Liwei Wang}, {and} \bibinfo{person}{Tie{-}Yan Liu}.} \bibinfo{year}{2019}\natexlab{}.
\newblock \showarticletitle{Representation Degeneration Problem in Training Natural Language Generation Models}. In \bibinfo{booktitle}{\emph{7th International Conference on Learning Representations, {ICLR} 2019, New Orleans, LA, USA, May 6-9, 2019}}. \bibinfo{publisher}{OpenReview.net}.
\newblock


\bibitem[Gao et~al\mbox{.}(2024)]%
        {gao2021clip}
\bibfield{author}{\bibinfo{person}{Peng Gao}, \bibinfo{person}{Shijie Geng}, \bibinfo{person}{Renrui Zhang}, \bibinfo{person}{Teli Ma}, \bibinfo{person}{Rongyao Fang}, \bibinfo{person}{Yongfeng Zhang}, \bibinfo{person}{Hongsheng Li}, {and} \bibinfo{person}{Yu Qiao}.} \bibinfo{year}{2024}\natexlab{}.
\newblock \showarticletitle{CLIP-Adapter: Better Vision-Language Models with Feature Adapters}.
\newblock \bibinfo{journal}{\emph{Int. J. Comput. Vis.}} \bibinfo{volume}{132}, \bibinfo{number}{2} (\bibinfo{year}{2024}), \bibinfo{pages}{581--595}.
\newblock
\urldef\tempurl%
\url{https://doi.org/10.1007/S11263-023-01891-X}
\showDOI{\tempurl}


\bibitem[Gao et~al\mbox{.}(2021)]%
        {gao2022simcse}
\bibfield{author}{\bibinfo{person}{Tianyu Gao}, \bibinfo{person}{Xingcheng Yao}, {and} \bibinfo{person}{Danqi Chen}.} \bibinfo{year}{2021}\natexlab{}.
\newblock \showarticletitle{SimCSE: Simple Contrastive Learning of Sentence Embeddings}. In \bibinfo{booktitle}{\emph{Proceedings of the 2021 Conference on Empirical Methods in Natural Language Processing, {EMNLP} 2021, Virtual Event / Punta Cana, Dominican Republic, 7-11 November, 2021}}. \bibinfo{publisher}{Association for Computational Linguistics}, \bibinfo{pages}{6894--6910}.
\newblock
\urldef\tempurl%
\url{https://doi.org/10.18653/V1/2021.EMNLP-MAIN.552}
\showDOI{\tempurl}


\bibitem[He et~al\mbox{.}(2014)]%
        {he2014predicting}
\bibfield{author}{\bibinfo{person}{Xiangnan He}, \bibinfo{person}{Ming Gao}, \bibinfo{person}{Min{-}Yen Kan}, \bibinfo{person}{Yiqun Liu}, {and} \bibinfo{person}{Kazunari Sugiyama}.} \bibinfo{year}{2014}\natexlab{}.
\newblock \showarticletitle{Predicting the popularity of web 2.0 items based on user comments}. In \bibinfo{booktitle}{\emph{The 37th International {ACM} {SIGIR} Conference on Research and Development in Information Retrieval, {SIGIR} '14, Gold Coast , QLD, Australia - July 06 - 11, 2014}}. \bibinfo{publisher}{{ACM}}, \bibinfo{pages}{233--242}.
\newblock
\urldef\tempurl%
\url{https://doi.org/10.1145/2600428.2609558}
\showDOI{\tempurl}


\bibitem[He et~al\mbox{.}(2019)]%
        {he2019feature}
\bibfield{author}{\bibinfo{person}{Ziliang He}, \bibinfo{person}{Zijian He}, \bibinfo{person}{Jiahong Wu}, {and} \bibinfo{person}{Zhenguo Yang}.} \bibinfo{year}{2019}\natexlab{}.
\newblock \showarticletitle{Feature Construction for Posts and Users Combined with LightGBM for Social Media Popularity Prediction}. In \bibinfo{booktitle}{\emph{Proceedings of the 27th {ACM} International Conference on Multimedia, {MM} 2019, Nice, France, October 21-25, 2019}}. \bibinfo{publisher}{{ACM}}, \bibinfo{pages}{2672--2676}.
\newblock
\urldef\tempurl%
\url{https://doi.org/10.1145/3343031.3356054}
\showDOI{\tempurl}


\bibitem[Hsu et~al\mbox{.}(2023)]%
        {Hsu2023}
\bibfield{author}{\bibinfo{person}{Chih{-}Chung Hsu}, \bibinfo{person}{Chia{-}Ming Lee}, \bibinfo{person}{Xiu{-}Yu Hou}, {and} \bibinfo{person}{Chi{-}Han Tsai}.} \bibinfo{year}{2023}\natexlab{}.
\newblock \showarticletitle{Gradient Boost Tree Network based on Extensive Feature Analysis for Popularity Prediction of Social Posts}. In \bibinfo{booktitle}{\emph{Proceedings of the 31st {ACM} International Conference on Multimedia, {MM} 2023, Ottawa, ON, Canada, 29 October 2023- 3 November 2023}}. \bibinfo{publisher}{{ACM}}, \bibinfo{pages}{9451--9455}.
\newblock
\urldef\tempurl%
\url{https://doi.org/10.1145/3581783.3612843}
\showDOI{\tempurl}


\bibitem[Ji et~al\mbox{.}(2023)]%
        {10.1145/3580305.3599281}
\bibfield{author}{\bibinfo{person}{Shuo Ji}, \bibinfo{person}{Xiaodong Lu}, \bibinfo{person}{Mingzhe Liu}, \bibinfo{person}{Leilei Sun}, \bibinfo{person}{Chuanren Liu}, \bibinfo{person}{Bowen Du}, {and} \bibinfo{person}{Hui Xiong}.} \bibinfo{year}{2023}\natexlab{}.
\newblock \showarticletitle{Community-Based Dynamic Graph Learning for Popularity Prediction}. In \bibinfo{booktitle}{\emph{Proceedings of the 29th ACM SIGKDD Conference on Knowledge Discovery and Data Mining}}. \bibinfo{pages}{930–940}.
\newblock


\bibitem[Kang et~al\mbox{.}(2019)]%
        {kang2019catboost}
\bibfield{author}{\bibinfo{person}{Peipei Kang}, \bibinfo{person}{Zehang Lin}, \bibinfo{person}{Shaohua Teng}, \bibinfo{person}{Guipeng Zhang}, \bibinfo{person}{Lingni Guo}, {and} \bibinfo{person}{Wei Zhang}.} \bibinfo{year}{2019}\natexlab{}.
\newblock \showarticletitle{Catboost-based Framework with Additional User Information for Social Media Popularity Prediction}. In \bibinfo{booktitle}{\emph{Proceedings of the 27th {ACM} International Conference on Multimedia, {MM} 2019, Nice, France, October 21-25, 2019}}. \bibinfo{publisher}{{ACM}}, \bibinfo{pages}{2677--2681}.
\newblock
\urldef\tempurl%
\url{https://doi.org/10.1145/3343031.3356060}
\showDOI{\tempurl}


\bibitem[Khosla et~al\mbox{.}(2014)]%
        {khosla2014makes}
\bibfield{author}{\bibinfo{person}{Aditya Khosla}, \bibinfo{person}{Atish~Das Sarma}, {and} \bibinfo{person}{Raffay Hamid}.} \bibinfo{year}{2014}\natexlab{}.
\newblock \showarticletitle{What makes an image popular?}. In \bibinfo{booktitle}{\emph{23rd International World Wide Web Conference, {WWW} '14, Seoul, Republic of Korea, April 7-11, 2014}}. \bibinfo{publisher}{{ACM}}, \bibinfo{pages}{867--876}.
\newblock
\urldef\tempurl%
\url{https://doi.org/10.1145/2566486.2567996}
\showDOI{\tempurl}


\bibitem[Khosla et~al\mbox{.}(2020)]%
        {khosla2021supervised}
\bibfield{author}{\bibinfo{person}{Prannay Khosla}, \bibinfo{person}{Piotr Teterwak}, \bibinfo{person}{Chen Wang}, \bibinfo{person}{Aaron Sarna}, \bibinfo{person}{Yonglong Tian}, \bibinfo{person}{Phillip Isola}, \bibinfo{person}{Aaron Maschinot}, \bibinfo{person}{Ce Liu}, {and} \bibinfo{person}{Dilip Krishnan}.} \bibinfo{year}{2020}\natexlab{}.
\newblock \showarticletitle{Supervised Contrastive Learning}. In \bibinfo{booktitle}{\emph{Advances in Neural Information Processing Systems 33: Annual Conference on Neural Information Processing Systems 2020, NeurIPS 2020, December 6-12, 2020, virtual}}.
\newblock


\bibitem[Lai et~al\mbox{.}(2020)]%
        {lai2020hyfea}
\bibfield{author}{\bibinfo{person}{Xin Lai}, \bibinfo{person}{Yihong Zhang}, {and} \bibinfo{person}{Wei Zhang}.} \bibinfo{year}{2020}\natexlab{}.
\newblock \showarticletitle{HyFea: Winning Solution to Social Media Popularity Prediction for Multimedia Grand Challenge 2020}. In \bibinfo{booktitle}{\emph{{MM} '20: The 28th {ACM} International Conference on Multimedia, Virtual Event / Seattle, WA, USA, October 12-16, 2020}}. \bibinfo{publisher}{{ACM}}, \bibinfo{pages}{4565--4569}.
\newblock
\urldef\tempurl%
\url{https://doi.org/10.1145/3394171.3416273}
\showDOI{\tempurl}


\bibitem[Li et~al\mbox{.}(2017)]%
        {li2017deepcas}
\bibfield{author}{\bibinfo{person}{Cheng Li}, \bibinfo{person}{Jiaqi Ma}, \bibinfo{person}{Xiaoxiao Guo}, {and} \bibinfo{person}{Qiaozhu Mei}.} \bibinfo{year}{2017}\natexlab{}.
\newblock \showarticletitle{DeepCas: An End-to-End Predictor of Information Cascades}. In \bibinfo{booktitle}{\emph{Proceedings of the 26th International Conference on World Wide Web}}. \bibinfo{pages}{577–586}.
\newblock


\bibitem[Li et~al\mbox{.}(2023)]%
        {li2023blip2}
\bibfield{author}{\bibinfo{person}{Junnan Li}, \bibinfo{person}{Dongxu Li}, \bibinfo{person}{Silvio Savarese}, {and} \bibinfo{person}{Steven Hoi}.} \bibinfo{year}{2023}\natexlab{}.
\newblock \bibinfo{title}{BLIP-2: Bootstrapping Language-Image Pre-training with Frozen Image Encoders and Large Language Models}.
\newblock
\newblock
\showeprint[arxiv]{2301.12597}~[cs.CV]


\bibitem[Liu et~al\mbox{.}(2023)]%
        {LIU2023103738}
\bibfield{author}{\bibinfo{person}{An{-}An Liu}, \bibinfo{person}{Hongwei Du}, \bibinfo{person}{Ning Xu}, \bibinfo{person}{Quan Zhang}, \bibinfo{person}{Shenyuan Zhang}, \bibinfo{person}{Yejun Tang}, {and} \bibinfo{person}{Xuanya Li}.} \bibinfo{year}{2023}\natexlab{}.
\newblock \showarticletitle{Exploring visual relationship for social media popularity prediction}.
\newblock \bibinfo{journal}{\emph{J. Vis. Commun. Image Represent.}}  \bibinfo{volume}{90} (\bibinfo{year}{2023}), \bibinfo{pages}{103738}.
\newblock
\urldef\tempurl%
\url{https://doi.org/10.1016/J.JVCIR.2022.103738}
\showDOI{\tempurl}


\bibitem[Liu et~al\mbox{.}(2024)]%
        {liu2024improved}
\bibfield{author}{\bibinfo{person}{Haotian Liu}, \bibinfo{person}{Chunyuan Li}, \bibinfo{person}{Yuheng Li}, {and} \bibinfo{person}{Yong~Jae Lee}.} \bibinfo{year}{2024}\natexlab{}.
\newblock \bibinfo{title}{Improved Baselines with Visual Instruction Tuning}.
\newblock
\newblock
\showeprint[arxiv]{2310.03744}~[cs.CV]


\bibitem[Manmatha et~al\mbox{.}(2017)]%
        {wu2018sampling}
\bibfield{author}{\bibinfo{person}{R. Manmatha}, \bibinfo{person}{Chao{-}Yuan Wu}, \bibinfo{person}{Alexander~J. Smola}, {and} \bibinfo{person}{Philipp Kr{\"{a}}henb{\"{u}}hl}.} \bibinfo{year}{2017}\natexlab{}.
\newblock \showarticletitle{Sampling Matters in Deep Embedding Learning}. In \bibinfo{booktitle}{\emph{{IEEE} International Conference on Computer Vision, {ICCV} 2017, Venice, Italy, October 22-29, 2017}}. \bibinfo{publisher}{{IEEE} Computer Society}, \bibinfo{pages}{2859--2867}.
\newblock
\urldef\tempurl%
\url{https://doi.org/10.1109/ICCV.2017.309}
\showDOI{\tempurl}


\bibitem[Pinto et~al\mbox{.}(2013)]%
        {pinto2013using}
\bibfield{author}{\bibinfo{person}{Henrique Pinto}, \bibinfo{person}{Jussara~M. Almeida}, {and} \bibinfo{person}{Marcos~Andr{\'{e}} Gon{\c{c}}alves}.} \bibinfo{year}{2013}\natexlab{}.
\newblock \showarticletitle{Using early view patterns to predict the popularity of youtube videos}. In \bibinfo{booktitle}{\emph{Sixth {ACM} International Conference on Web Search and Data Mining, {WSDM} 2013, Rome, Italy, February 4-8, 2013}}. \bibinfo{publisher}{{ACM}}, \bibinfo{pages}{365--374}.
\newblock
\urldef\tempurl%
\url{https://doi.org/10.1145/2433396.2433443}
\showDOI{\tempurl}


\bibitem[Qiu et~al\mbox{.}(2022)]%
        {Qiu_2022}
\bibfield{author}{\bibinfo{person}{Ruihong Qiu}, \bibinfo{person}{Zi Huang}, \bibinfo{person}{Hongzhi Yin}, {and} \bibinfo{person}{Zijian Wang}.} \bibinfo{year}{2022}\natexlab{}.
\newblock \showarticletitle{Contrastive Learning for Representation Degeneration Problem in Sequential Recommendation}. In \bibinfo{booktitle}{\emph{{WSDM} '22: The Fifteenth {ACM} International Conference on Web Search and Data Mining, Virtual Event / Tempe, AZ, USA, February 21 - 25, 2022}}. \bibinfo{publisher}{{ACM}}, \bibinfo{pages}{813--823}.
\newblock
\urldef\tempurl%
\url{https://doi.org/10.1145/3488560.3498433}
\showDOI{\tempurl}


\bibitem[Radford et~al\mbox{.}(2021)]%
        {radford2021learning}
\bibfield{author}{\bibinfo{person}{Alec Radford}, \bibinfo{person}{Jong~Wook Kim}, \bibinfo{person}{Chris Hallacy}, \bibinfo{person}{Aditya Ramesh}, \bibinfo{person}{Gabriel Goh}, \bibinfo{person}{Sandhini Agarwal}, \bibinfo{person}{Girish Sastry}, \bibinfo{person}{Amanda Askell}, \bibinfo{person}{Pamela Mishkin}, \bibinfo{person}{Jack Clark}, \bibinfo{person}{Gretchen Krueger}, {and} \bibinfo{person}{Ilya Sutskever}.} \bibinfo{year}{2021}\natexlab{}.
\newblock \showarticletitle{Learning Transferable Visual Models From Natural Language Supervision}. In \bibinfo{booktitle}{\emph{Proceedings of the 38th International Conference on Machine Learning, {ICML} 2021, 18-24 July 2021, Virtual Event}} \emph{(\bibinfo{series}{Proceedings of Machine Learning Research}, Vol.~\bibinfo{volume}{139})}. \bibinfo{publisher}{{PMLR}}, \bibinfo{pages}{8748--8763}.
\newblock


\bibitem[Shahbaznezhad et~al\mbox{.}(2021)]%
        {doi:10.1016/j.intmar.2020.05.001}
\bibfield{author}{\bibinfo{person}{Hamidreza Shahbaznezhad}, \bibinfo{person}{Rebecca Dolan}, {and} \bibinfo{person}{Mona Rashidirad}.} \bibinfo{year}{2021}\natexlab{}.
\newblock \showarticletitle{The Role of Social Media Content Format and Platform in Users’ Engagement Behavior}.
\newblock \bibinfo{journal}{\emph{Journal of Interactive Marketing}} \bibinfo{volume}{53}, \bibinfo{number}{1} (\bibinfo{year}{2021}), \bibinfo{pages}{47--65}.
\newblock
\urldef\tempurl%
\url{https://doi.org/10.1016/j.intmar.2020.05.001}
\showDOI{\tempurl}


\bibitem[Stasi(2019)]%
        {doi:10.1177/1783591719847545}
\bibfield{author}{\bibinfo{person}{Maria~Luisa Stasi}.} \bibinfo{year}{2019}\natexlab{}.
\newblock \showarticletitle{Social media platforms and content exposure: How to restore users’ control}.
\newblock \bibinfo{journal}{\emph{Competition and Regulation in Network Industries}} \bibinfo{volume}{20}, \bibinfo{number}{1} (\bibinfo{year}{2019}), \bibinfo{pages}{86--110}.
\newblock
\urldef\tempurl%
\url{https://doi.org/10.1177/1783591719847545}
\showDOI{\tempurl}


\bibitem[Tan et~al\mbox{.}(2022)]%
        {tan2022efficient}
\bibfield{author}{\bibinfo{person}{Yunpeng Tan}, \bibinfo{person}{Fangyu Liu}, \bibinfo{person}{Bowei Li}, \bibinfo{person}{Zheng Zhang}, {and} \bibinfo{person}{Bo Zhang}.} \bibinfo{year}{2022}\natexlab{}.
\newblock \showarticletitle{An Efficient Multi-View Multimodal Data Processing Framework for Social Media Popularity Prediction}. In \bibinfo{booktitle}{\emph{{MM} '22: The 30th {ACM} International Conference on Multimedia, Lisboa, Portugal, October 10 - 14, 2022}}. \bibinfo{publisher}{{ACM}}, \bibinfo{pages}{7200--7204}.
\newblock
\urldef\tempurl%
\url{https://doi.org/10.1145/3503161.3551607}
\showDOI{\tempurl}


\bibitem[Wang et~al\mbox{.}(2023a)]%
        {wang2023deeper}
\bibfield{author}{\bibinfo{person}{Fangye Wang}, \bibinfo{person}{Hansu Gu}, \bibinfo{person}{Dongsheng Li}, \bibinfo{person}{Tun Lu}, \bibinfo{person}{Peng Zhang}, {and} \bibinfo{person}{Ning Gu}.} \bibinfo{year}{2023}\natexlab{a}.
\newblock \showarticletitle{Towards Deeper, Lighter and Interpretable Cross Network for {CTR} Prediction}. In \bibinfo{booktitle}{\emph{Proceedings of the 32nd {ACM} International Conference on Information and Knowledge Management, {CIKM} 2023, Birmingham, United Kingdom, October 21-25, 2023}}. \bibinfo{publisher}{{ACM}}, \bibinfo{pages}{2523--2533}.
\newblock
\urldef\tempurl%
\url{https://doi.org/10.1145/3583780.3615089}
\showDOI{\tempurl}


\bibitem[Wang et~al\mbox{.}(2023b)]%
        {WANG2023101490}
\bibfield{author}{\bibinfo{person}{Jing Wang}, \bibinfo{person}{Shuo Yang}, \bibinfo{person}{Hui Zhao}, {and} \bibinfo{person}{Yue Yang}.} \bibinfo{year}{2023}\natexlab{b}.
\newblock \showarticletitle{Social media popularity prediction with multimodal hierarchical fusion model}.
\newblock \bibinfo{journal}{\emph{Comput. Speech Lang.}}  \bibinfo{volume}{80} (\bibinfo{year}{2023}), \bibinfo{pages}{101490}.
\newblock
\urldef\tempurl%
\url{https://doi.org/10.1016/J.CSL.2023.101490}
\showDOI{\tempurl}


\bibitem[Wang et~al\mbox{.}(2020)]%
        {wang2020feature}
\bibfield{author}{\bibinfo{person}{Kai Wang}, \bibinfo{person}{Penghui Wang}, \bibinfo{person}{Xin Chen}, \bibinfo{person}{Qiushi Huang}, \bibinfo{person}{Zhendong Mao}, {and} \bibinfo{person}{Yongdong Zhang}.} \bibinfo{year}{2020}\natexlab{}.
\newblock \showarticletitle{A Feature Generalization Framework for Social Media Popularity Prediction}. In \bibinfo{booktitle}{\emph{{MM} '20: The 28th {ACM} International Conference on Multimedia, Virtual Event / Seattle, WA, USA, October 12-16, 2020}}. \bibinfo{publisher}{{ACM}}, \bibinfo{pages}{4570--4574}.
\newblock
\urldef\tempurl%
\url{https://doi.org/10.1145/3394171.3416294}
\showDOI{\tempurl}


\bibitem[Wang and Isola(2020)]%
        {wang2022contra}
\bibfield{author}{\bibinfo{person}{Tongzhou Wang} {and} \bibinfo{person}{Phillip Isola}.} \bibinfo{year}{2020}\natexlab{}.
\newblock \showarticletitle{Understanding Contrastive Representation Learning through Alignment and Uniformity on the Hypersphere}. In \bibinfo{booktitle}{\emph{Proceedings of the 37th International Conference on Machine Learning, {ICML} 2020, 13-18 July 2020, Virtual Event}} \emph{(\bibinfo{series}{Proceedings of Machine Learning Research}, Vol.~\bibinfo{volume}{119})}. \bibinfo{publisher}{{PMLR}}, \bibinfo{pages}{9929--9939}.
\newblock


\bibitem[Wang and Zhang(2017)]%
        {wang2017combining}
\bibfield{author}{\bibinfo{person}{Wen Wang} {and} \bibinfo{person}{Wei Zhang}.} \bibinfo{year}{2017}\natexlab{}.
\newblock \showarticletitle{Combining Multiple Features for Image Popularity Prediction in Social Media}. In \bibinfo{booktitle}{\emph{Proceedings of the 2017 {ACM} on Multimedia Conference, {MM} 2017, Mountain View, CA, USA, October 23-27, 2017}}. \bibinfo{publisher}{{ACM}}, \bibinfo{pages}{1901--1905}.
\newblock
\urldef\tempurl%
\url{https://doi.org/10.1145/3123266.3127900}
\showDOI{\tempurl}


\bibitem[Wu et~al\mbox{.}(2019)]%
        {wu2019smp}
\bibfield{author}{\bibinfo{person}{Bo Wu}, \bibinfo{person}{Wen{-}Huang Cheng}, \bibinfo{person}{Peiye Liu}, \bibinfo{person}{Bei Liu}, \bibinfo{person}{Zhaoyang Zeng}, {and} \bibinfo{person}{Jiebo Luo}.} \bibinfo{year}{2019}\natexlab{}.
\newblock \showarticletitle{{SMP} Challenge: An Overview of Social Media Prediction Challenge 2019}. In \bibinfo{booktitle}{\emph{Proceedings of the 27th {ACM} International Conference on Multimedia, {MM} 2019, Nice, France, October 21-25, 2019}}. \bibinfo{publisher}{{ACM}}, \bibinfo{pages}{2667--2671}.
\newblock
\urldef\tempurl%
\url{https://doi.org/10.1145/3343031.3356084}
\showDOI{\tempurl}


\bibitem[Wu et~al\mbox{.}(2017)]%
        {wu2017sequential}
\bibfield{author}{\bibinfo{person}{Bo Wu}, \bibinfo{person}{Wen{-}Huang Cheng}, \bibinfo{person}{Yongdong Zhang}, \bibinfo{person}{Qiushi Huang}, \bibinfo{person}{Jintao Li}, {and} \bibinfo{person}{Tao Mei}.} \bibinfo{year}{2017}\natexlab{}.
\newblock \showarticletitle{Sequential Prediction of Social Media Popularity with Deep Temporal Context Networks}. In \bibinfo{booktitle}{\emph{Proceedings of the Twenty-Sixth International Joint Conference on Artificial Intelligence, {IJCAI} 2017, Melbourne, Australia, August 19-25, 2017}}. \bibinfo{publisher}{ijcai.org}, \bibinfo{pages}{3062--3068}.
\newblock
\urldef\tempurl%
\url{https://doi.org/10.24963/IJCAI.2017/427}
\showDOI{\tempurl}


\bibitem[Wu et~al\mbox{.}(2022)]%
        {wu2022deeply}
\bibfield{author}{\bibinfo{person}{Jianmin Wu}, \bibinfo{person}{Liming Zhao}, \bibinfo{person}{Dangwei Li}, \bibinfo{person}{Chen{-}Wei Xie}, \bibinfo{person}{Siyang Sun}, {and} \bibinfo{person}{Yun Zheng}.} \bibinfo{year}{2022}\natexlab{}.
\newblock \showarticletitle{Deeply Exploit Visual and Language Information for Social Media Popularity Prediction}. In \bibinfo{booktitle}{\emph{{MM} '22: The 30th {ACM} International Conference on Multimedia, Lisboa, Portugal, October 10 - 14, 2022}}. \bibinfo{publisher}{{ACM}}, \bibinfo{pages}{7045--7049}.
\newblock
\urldef\tempurl%
\url{https://doi.org/10.1145/3503161.3551576}
\showDOI{\tempurl}


\bibitem[Wu et~al\mbox{.}(2020)]%
        {wu2020clear}
\bibfield{author}{\bibinfo{person}{Zhuofeng Wu}, \bibinfo{person}{Sinong Wang}, \bibinfo{person}{Jiatao Gu}, \bibinfo{person}{Madian Khabsa}, \bibinfo{person}{Fei Sun}, {and} \bibinfo{person}{Hao Ma}.} \bibinfo{year}{2020}\natexlab{}.
\newblock \showarticletitle{{CLEAR:} Contrastive Learning for Sentence Representation}.
\newblock \bibinfo{journal}{\emph{CoRR}}  \bibinfo{volume}{abs/2012.15466} (\bibinfo{year}{2020}).
\newblock
\showeprint[arXiv]{2012.15466}


\bibitem[Xu et~al\mbox{.}(2020)]%
        {xu2020multimodal}
\bibfield{author}{\bibinfo{person}{Kele Xu}, \bibinfo{person}{Zhimin Lin}, \bibinfo{person}{Jianqiao Zhao}, \bibinfo{person}{Peichang Shi}, \bibinfo{person}{Wei Deng}, {and} \bibinfo{person}{Huaimin Wang}.} \bibinfo{year}{2020}\natexlab{}.
\newblock \showarticletitle{Multimodal Deep Learning for Social Media Popularity Prediction With Attention Mechanism}. In \bibinfo{booktitle}{\emph{{MM} '20: The 28th {ACM} International Conference on Multimedia, Virtual Event / Seattle, WA, USA, October 12-16, 2020}}. \bibinfo{publisher}{{ACM}}, \bibinfo{pages}{4580--4584}.
\newblock
\urldef\tempurl%
\url{https://doi.org/10.1145/3394171.3416274}
\showDOI{\tempurl}


\bibitem[Yu et~al\mbox{.}(2021)]%
        {yu2021xcrossnet}
\bibfield{author}{\bibinfo{person}{Runlong Yu}, \bibinfo{person}{Yuyang Ye}, \bibinfo{person}{Qi Liu}, \bibinfo{person}{Zihan Wang}, \bibinfo{person}{Chunfeng Yang}, \bibinfo{person}{Yucheng Hu}, {and} \bibinfo{person}{Enhong Chen}.} \bibinfo{year}{2021}\natexlab{}.
\newblock \showarticletitle{XCrossNet: Feature Structure-Oriented Learning for Click-Through Rate Prediction}. In \bibinfo{booktitle}{\emph{Advances in Knowledge Discovery and Data Mining - 25th Pacific-Asia Conference, {PAKDD} 2021, Virtual Event, May 11-14, 2021, Proceedings, Part {II}}} \emph{(\bibinfo{series}{Lecture Notes in Computer Science}, Vol.~\bibinfo{volume}{12713})}. \bibinfo{publisher}{Springer}, \bibinfo{pages}{436--447}.
\newblock
\urldef\tempurl%
\url{https://doi.org/10.1007/978-3-030-75765-6\_35}
\showDOI{\tempurl}


\bibitem[Zha et~al\mbox{.}(2023)]%
        {zha2023rankncontrast}
\bibfield{author}{\bibinfo{person}{Kaiwen Zha}, \bibinfo{person}{Peng Cao}, \bibinfo{person}{Jeany Son}, \bibinfo{person}{Yuzhe Yang}, {and} \bibinfo{person}{Dina Katabi}.} \bibinfo{year}{2023}\natexlab{}.
\newblock \showarticletitle{Rank-N-Contrast: Learning Continuous Representations for Regression}. In \bibinfo{booktitle}{\emph{Advances in Neural Information Processing Systems 36: Annual Conference on Neural Information Processing Systems 2023, NeurIPS 2023, New Orleans, LA, USA, December 10 - 16, 2023}}.
\newblock


\bibitem[Zhang et~al\mbox{.}(2022)]%
        {zhang2022use}
\bibfield{author}{\bibinfo{person}{Shu Zhang}, \bibinfo{person}{Ran Xu}, \bibinfo{person}{Caiming Xiong}, {and} \bibinfo{person}{Chetan Ramaiah}.} \bibinfo{year}{2022}\natexlab{}.
\newblock \showarticletitle{Use All The Labels: {A} Hierarchical Multi-Label Contrastive Learning Framework}. In \bibinfo{booktitle}{\emph{{IEEE/CVF} Conference on Computer Vision and Pattern Recognition, {CVPR} 2022, New Orleans, LA, USA, June 18-24, 2022}}. \bibinfo{publisher}{{IEEE}}, \bibinfo{pages}{16639--16648}.
\newblock
\urldef\tempurl%
\url{https://doi.org/10.1109/CVPR52688.2022.01616}
\showDOI{\tempurl}


\bibitem[Zhang et~al\mbox{.}(2023)]%
        {zhang2023improving}
\bibfield{author}{\bibinfo{person}{Zhizhen Zhang}, \bibinfo{person}{Xiaohui Xie}, \bibinfo{person}{Mengyu Yang}, \bibinfo{person}{Ye Tian}, \bibinfo{person}{Yong Jiang}, {and} \bibinfo{person}{Yong Cui}.} \bibinfo{year}{2023}\natexlab{}.
\newblock \showarticletitle{Improving Social Media Popularity Prediction with Multiple Post Dependencies}.
\newblock \bibinfo{journal}{\emph{CoRR}}  \bibinfo{volume}{abs/2307.15413} (\bibinfo{year}{2023}).
\newblock
\urldef\tempurl%
\url{https://doi.org/10.48550/ARXIV.2307.15413}
\showDOI{\tempurl}
\showeprint[arXiv]{2307.15413}


\end{thebibliography}

\setcounter{table}{0}   
\setcounter{figure}{0}
\renewcommand{\thetable}{A\arabic{table}}
\renewcommand{\thefigure}{A\arabic{figure}}
\begin{appendix}

\section{Data Description}\label{app:user}
We provide the statistical information of three datasets. From Table \ref{tab:datasets}, we can see as the size of the dataset decreases, the average number of posts per user also decreases, coinciding with the ablation experiments in which the UID task works better on large datasets. At the same time, the average number of posts contained in each level in the category label also decreases, increasing the difficulty of learning.\par
All the user information we use is listed in Table \ref{tab:metadata}. For categorical data $S$, we adopt learnable embedding layers to transform them into features $E$. For numerical data $D$, we first transform them by super-linear and sub-linear operations, i.e., $D^2$ and $\sqrt{D}$, and we concatenate them to get augmented numerical data $\tilde{D}=[D,D^2,\sqrt{D}]$. Then we normalize $\tilde{D}$ by min-max normalization.
\begin{table}[htb]
    \centering
    \begin{tabular}{lc}
    \toprule
    & Data Entries \\
    \midrule
    \multirow{3}{*}{Categorical}
         &  user\_id; ispro; canbuy\_pro; \\
         & ispublic; timezone\_id; timezone\_offset; \\
         & post\_hour; post\_day; post\_month \\
    \midrule
    \multirow{4}{*}{Numerical}
    & totalViews; totalTags; totalGeotagged; \\
    & totalFaves; totalInGroup; photoCount; \\
    & followerCount; followingCount; geo\_acc;\\
    & pfirst\_year; pfirst\_taken\_year \\
    \bottomrule
    \end{tabular}
    \caption{The user information used for prediction.}
    \label{tab:metadata}
\end{table}

\section{Details of Baselines}\label{app:baseline}
Of the seven popularity prediction baselines we compared, only DTCN\footnote{https://github.com/csbobby/DTCN\_IJCAI} and DSN\footnote{https://github.com/Daisy-zzz/DSN} released their codes, so we reproduced the other five baselines according to their papers. We choose the hyperparameters according to the experimental performance in the original papers. For DTCN, the time unit $t_{\text {unit }}$ and the temporal range of context $t_r$ are chosen to be $1$ day. For Multiview, the time window size is set to $5$. For TTC-VLT, the number of self-attention layers is $12$ and the number of heads in each layer is $12$. The authors combine TTC-VLT and another catboost model to get the best result, but we only use TTC-VLT for a fair comparison. For DSN, the length of the post sequence is set to $8$, and the residual ratios are set to $\alpha=0.2$ and $\beta=0.6$. For VisualR, the number of relationships $M$ is chosen as $10$, the similarity threshold $\sigma$ is $0.8$, and the times threshold $\gamma$ is $5$. For other parameters that were not experimented with in the original paper, we directly use the values given in the paper. For the contrastive regression baseline RNC\footnote{https://github.com/kaiwenzha/Rank-N-Contrast}, we add it to our base model PP as a supervised contrastive loss:
\begin{equation}
    \mathcal{L}_{RNC} = \sum_{i \in I} \frac{-1}{|A(i)|} \sum_{j \in A(i)} log \frac{e^{\operatorname{sim}\left(\mathbf{f}_i^{pop}, \mathbf{f}_j^{pop}\right) / \tau}}{\sum_{a \in \mathcal{S}_{i,j}} e^{\operatorname{sim}\left(\mathbf{f}_i^{pop}, \mathbf{f}_a^{pop}\right) / \tau}} 
\end{equation}
where $\mathcal{S}_{i,j}={\left\{f_k \mid k \neq i, d\left(\hat{y}_i, \hat{y}_k\right) \geq d\left(\hat{y}_i, \hat{y}_j\right)\right\}}$ denoting the set of samples that are of higher ranks than $f_j$ in terms of popularity label distance, $f_i^{pop}$ is the final output of sample $x_i$ through $Enc_{pop}$ computed by Eq. \ref{eq:encpop}, $\tau$ is tuned to be $2$. We let the gradient be $\nabla \mathcal{L}_{RNC}=\frac{\partial \operatorname{\mathcal{L}_{RNC}}}{\partial \Theta}$, which can flow though all parameters of the model. The total loss then can be:
\begin{equation}
    \mathcal{L} = \lambda \mathcal{L}_{reg} + (1-\lambda)\mathcal{L}_{RNC} 
\end{equation}
where $\lambda$ is chosen as $0.9$ as the same of our model. 
All methods are tuned to the best performance with early stopping when validation errors have not declined for 10 consecutive epochs.

\begin{figure}[t]
\begin{subfigure}{0.23\textwidth}
    \includegraphics[width=\linewidth]{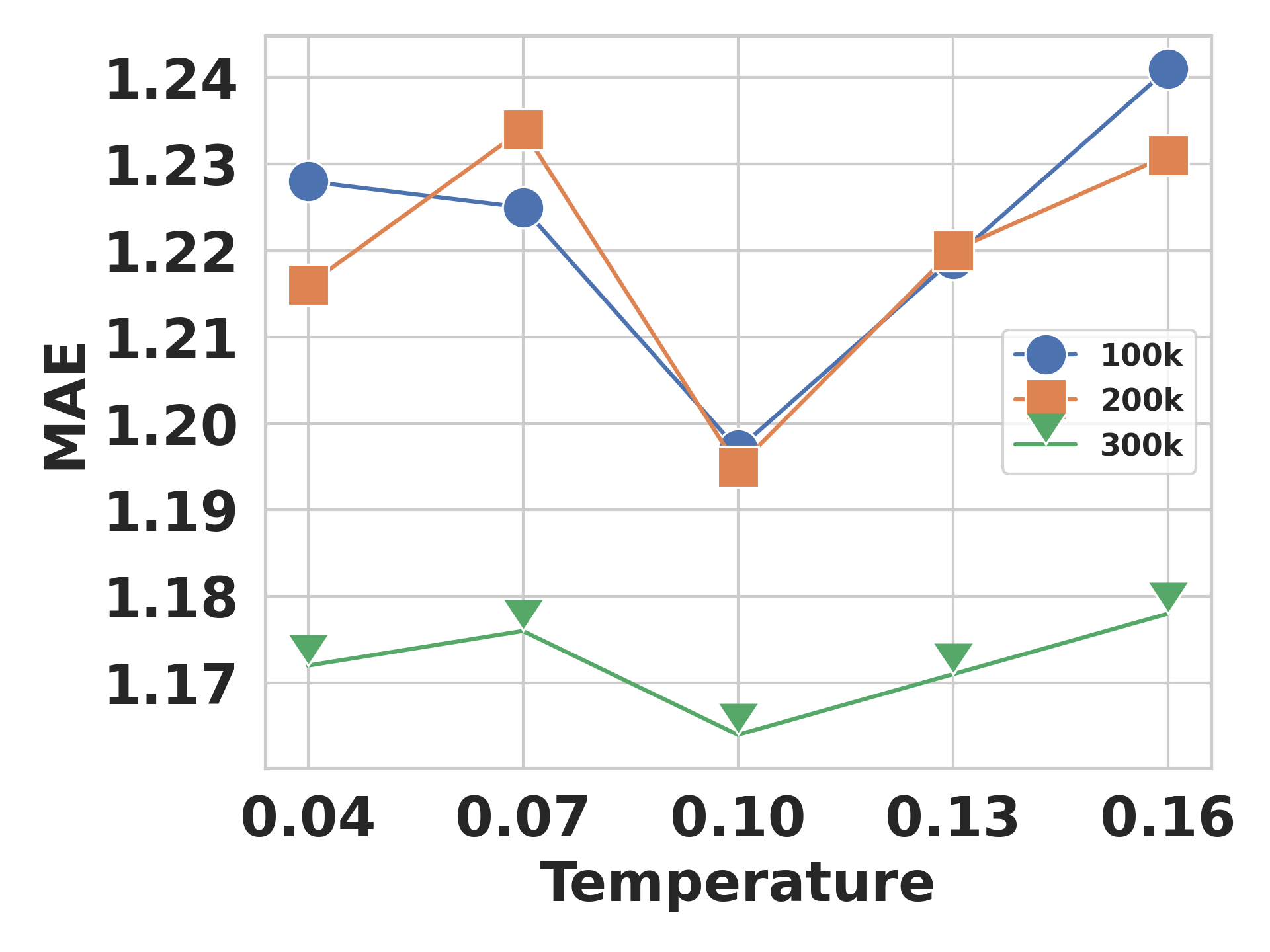}
    \caption{}
    \label{fig:param-temp}
  \end{subfigure}
  \hfill
  \begin{subfigure}{0.23\textwidth}
    \includegraphics[width=\linewidth]{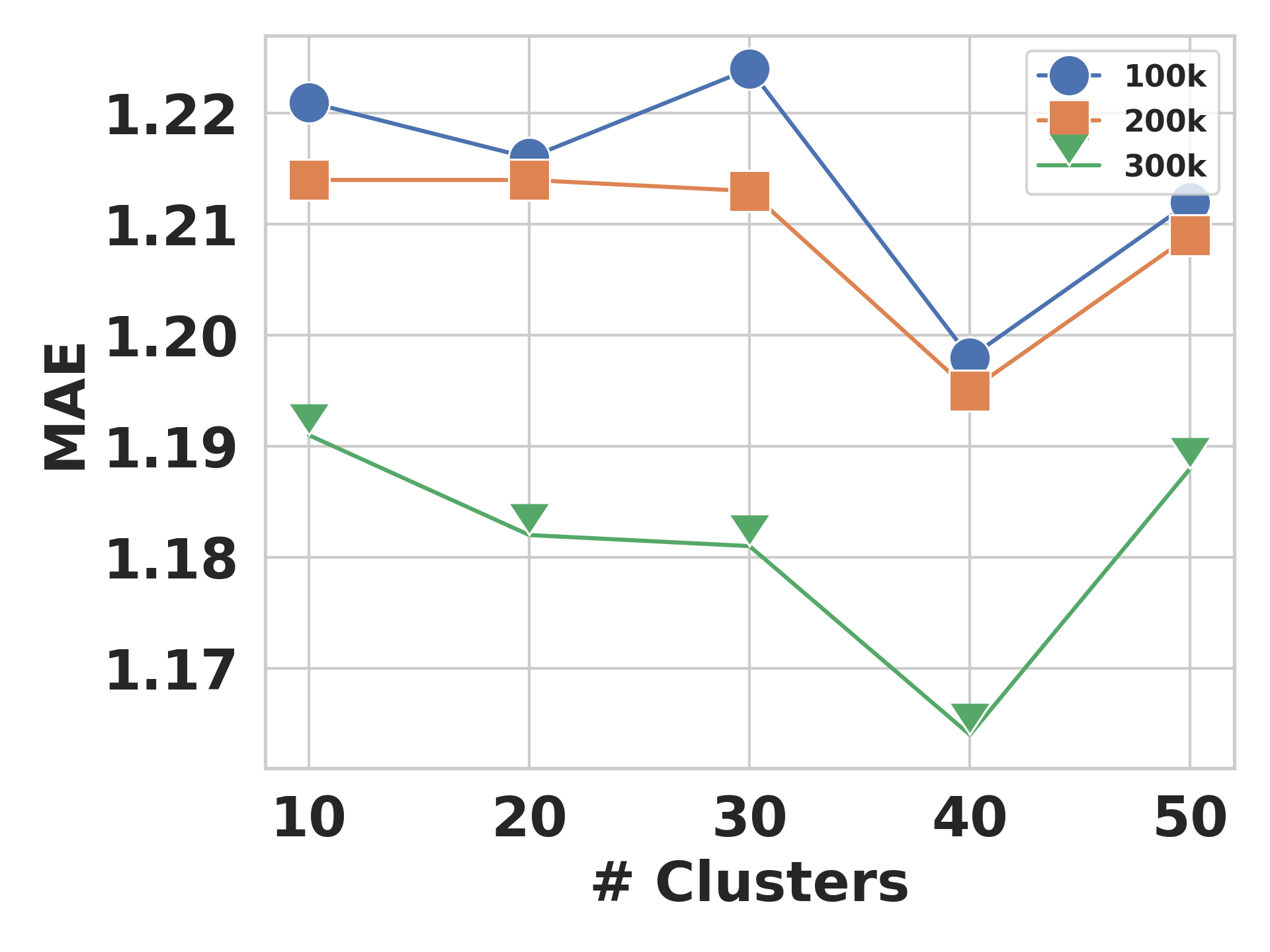}
    \caption{}
    \label{fig:param-cluster}
  \end{subfigure}
  \hfill
  \begin{subfigure}{0.23\textwidth}
    \includegraphics[width=\linewidth]{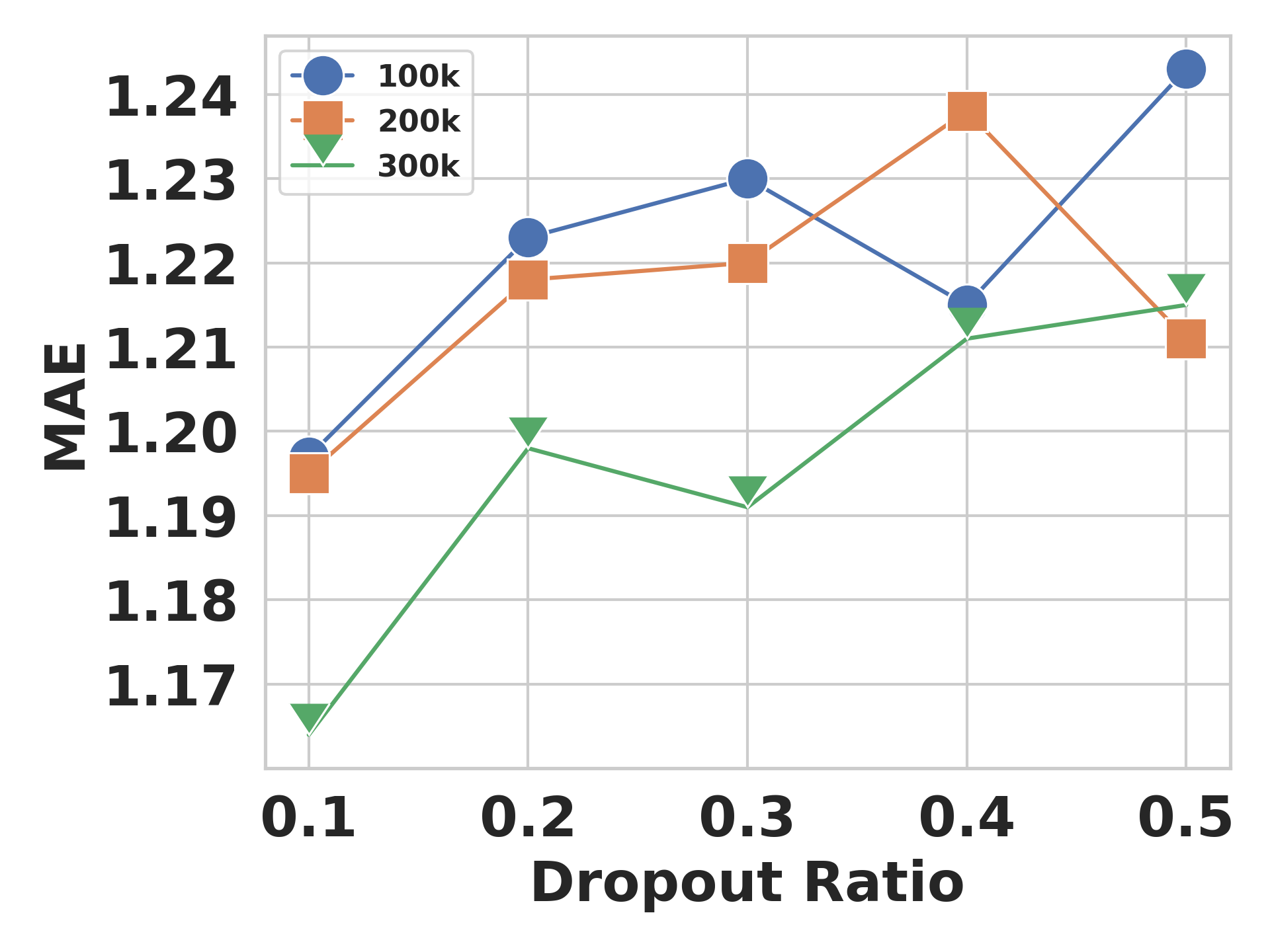}
    \caption{}
    \label{fig:param-drop}
  \end{subfigure}
  \hfill
  \begin{subfigure}{0.23\textwidth}
    \includegraphics[width=\linewidth]{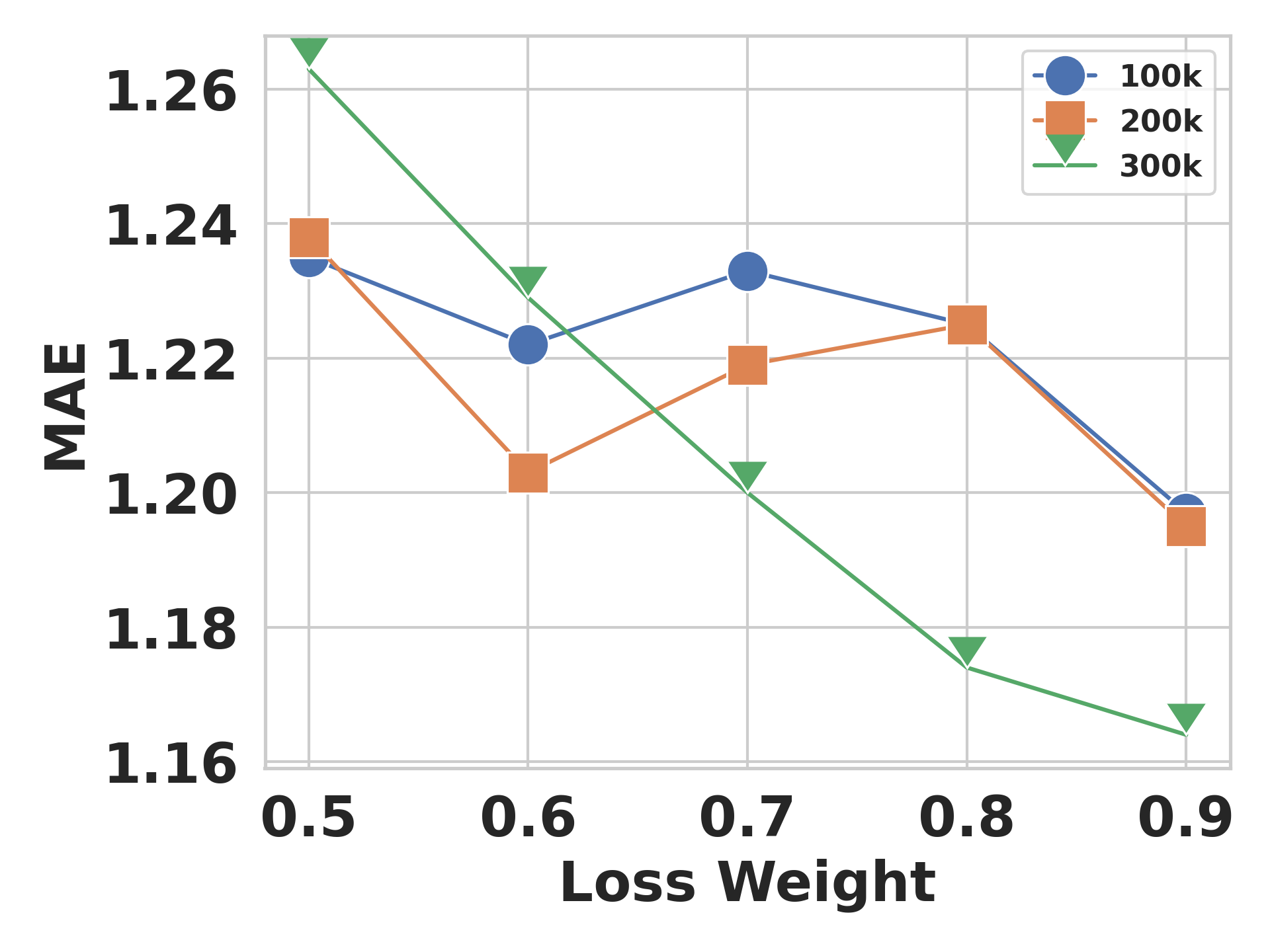}
    \caption{}
    \label{fig:param-loss}
  \end{subfigure}
  \label{fig:param}
  \begin{subfigure}{0.23\textwidth}
    \includegraphics[width=\linewidth]{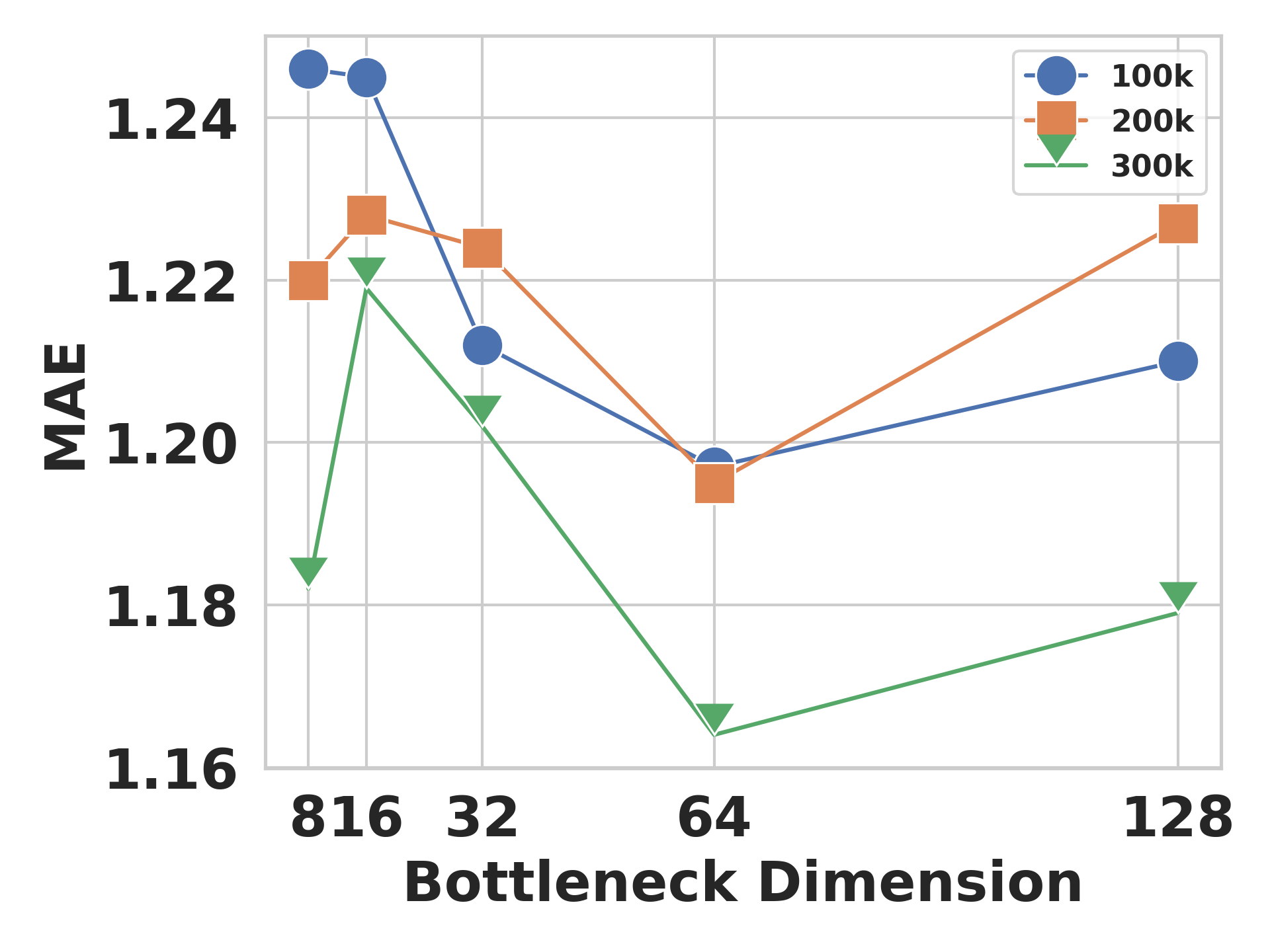}
    \caption{}
    \label{fig:param-neck}
  \end{subfigure}
  \hfill
  \begin{subfigure}{0.23\textwidth}
    \includegraphics[width=\linewidth]{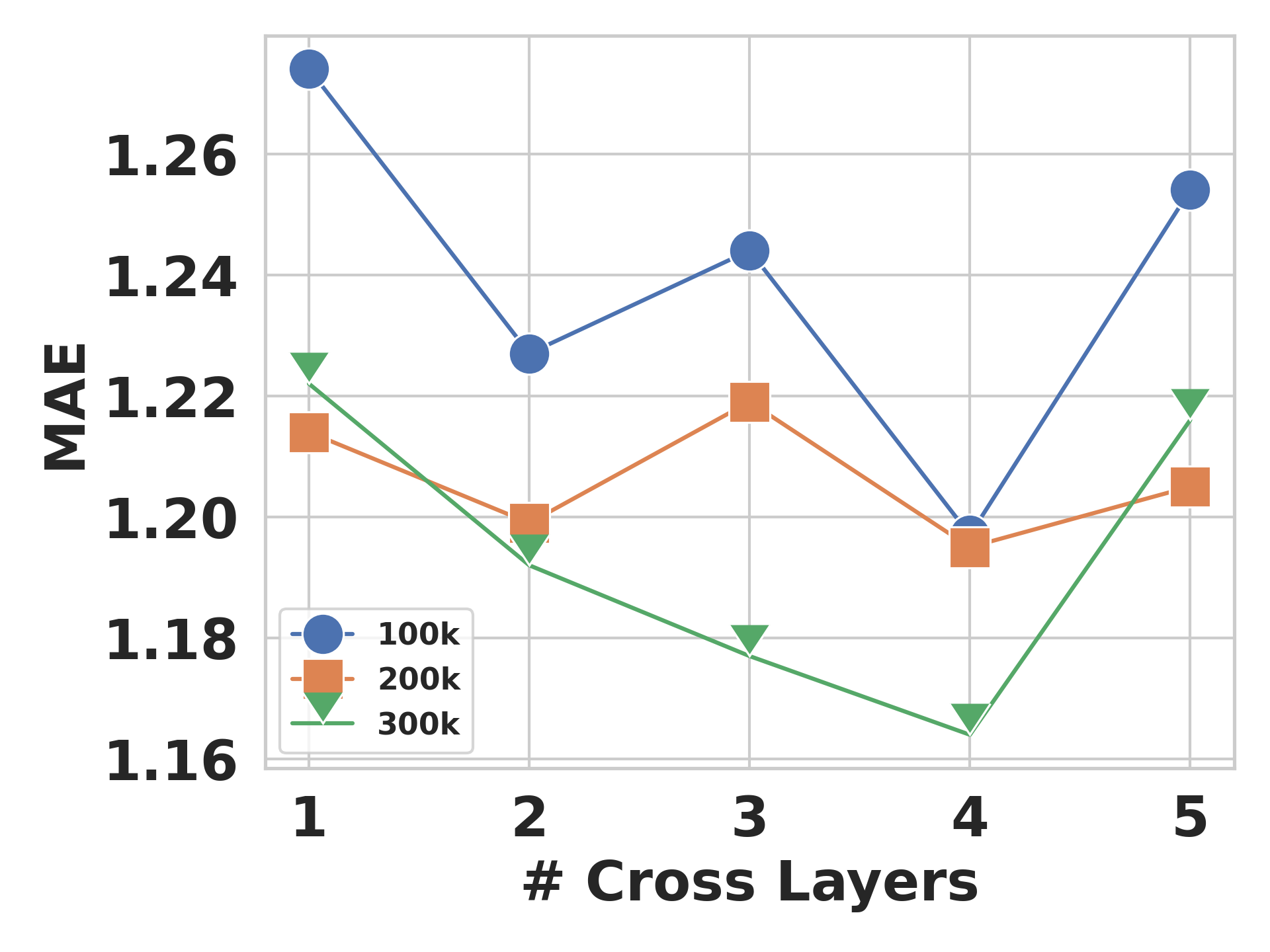}
    \caption{}
    \label{fig:param-cross}
  \end{subfigure}
  \hfill
  \caption{Parameter sensitivity analysis.}
  \label{fig:param}
\end{figure}

\begin{table*}[tb]
    \centering
    \begin{tabular}{l|c|c|c}
    \toprule
         & SMPD-$100K$ & SMPD-$200K$ & SMPD-$300K$ \\
    \midrule
        $\#$post & $101,865$ & $203,730$ & $305,595$ \\
        $\#$user & $21,217$ & $31,093$ & $38,307$ \\
        $\#$category & $11/77/665$ & $11/77/667$ & $11/77/668$ \\
        Avg. $\#$post / user & $4.8$ & $6.6$ & $8.0$ \\
        Avg. $\#$post / category & $9,260/1323/153$ & $18,521/2,646/305$ & $27,781/3,969/457$ \\
        Avg. popularity & $6.413$ & $6.411$ & $6.406$ \\
    \bottomrule
    \end{tabular}
    \caption{Dataset Statistics.}
    \label{tab:datasets}
\end{table*}

\begin{table}[tb]
    \centering
    \begin{tabular}{lcccccc}
    \toprule \multirow{2}{*}{ Method } & \multicolumn{2}{c}{ SMPD-$100K$ } & \multicolumn{2}{c}{ SMPD-$200K$ } & \multicolumn{2}{c}{ SMPD-$300K$ } \\
        \cline { 2 - 7 } & MAE & SRC & MAE & SRC & MAE & SRC \\
        \midrule
        CLIP  & 1.197 & 0.737 & 1.195 & 0.741 & 1.164 & 0.762 \\
        ResNet+BERT & 1.279 & 0.699 & 1.244 & 0.717 & 1.208 & 0.727 \\
        Llava-1.5 7b & 1.236 & 0.726 & 1.229 & 0.737 & 1.193 & 0.760 \\

        \bottomrule
    \end{tabular}
    \caption{Sensitivity to pre-trained models.}
    \label{tab:vtfeat}
\end{table}

\begin{figure*}[tb]
    \centering
    \includegraphics[width=0.8\linewidth]{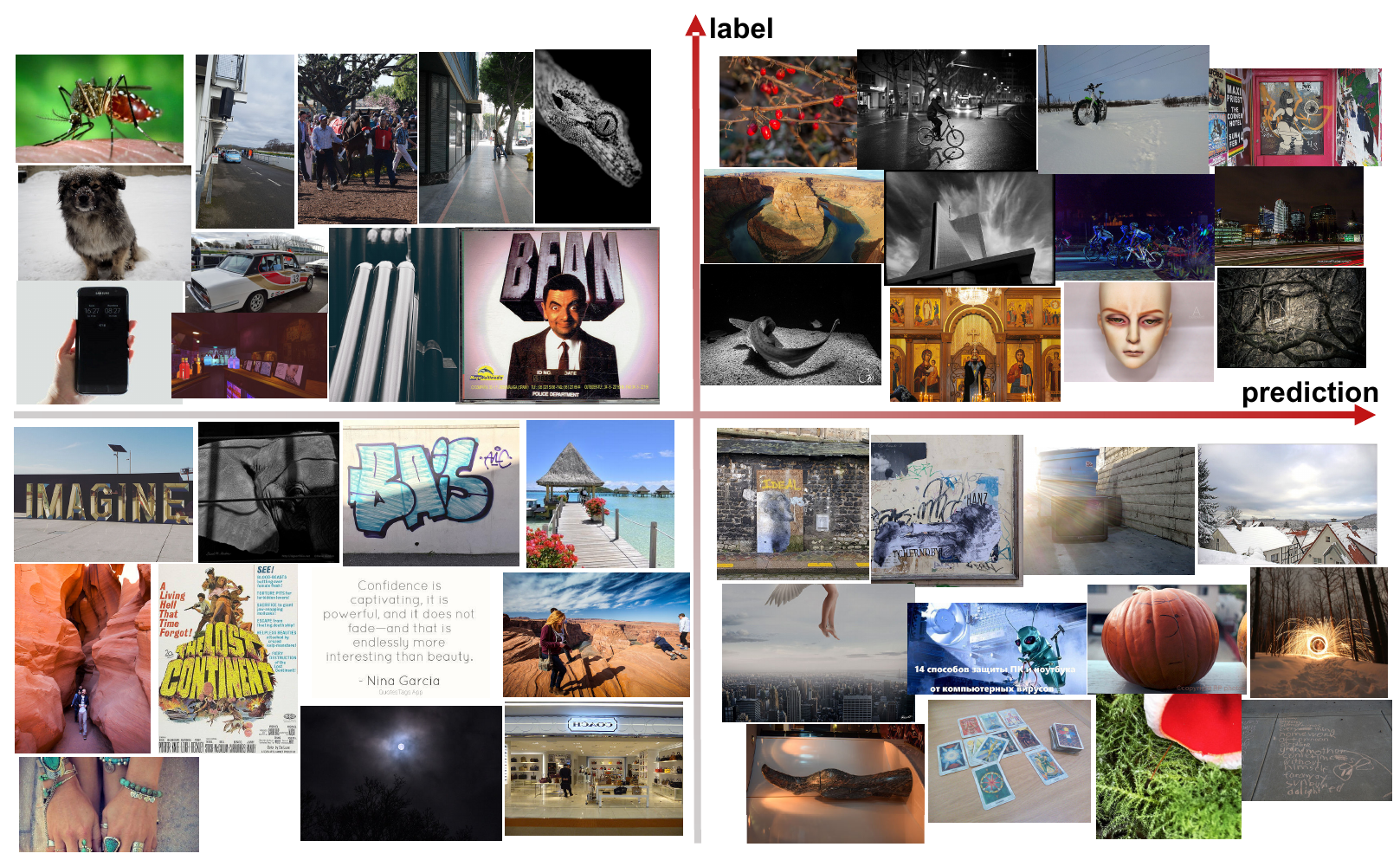}
    \caption{Visualization of post popularity. The left-to-right horizontal axis represents predicted popularity from small to large, and the bottom-to-top vertical axis represents popularity labels from small to large.}
    \label{fig:case}
\end{figure*}

\begin{figure}[tb]
  \begin{subfigure}{0.21\textwidth}
    \includegraphics[width=\linewidth]{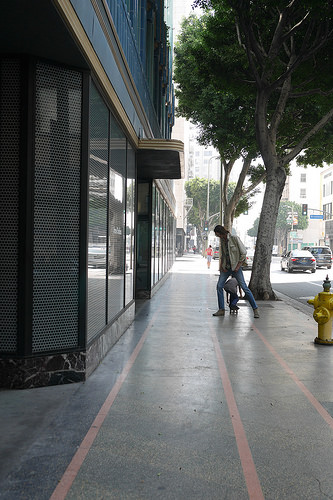}
    \caption{}
    \label{fig:case-1}
  \end{subfigure}
  \hfill
  \begin{subfigure}{0.21\textwidth}
    \includegraphics[width=\linewidth]{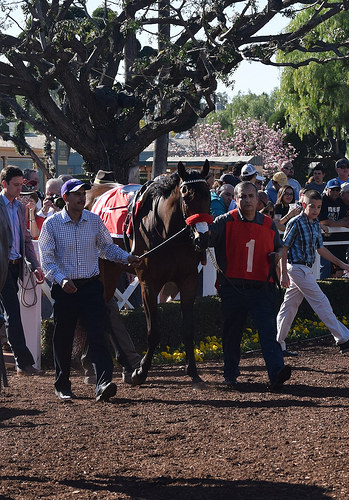}
    \caption{}
    \label{fig:case-2}
  \end{subfigure}
  \hfill
  \begin{subfigure}{0.21\textwidth}
    \includegraphics[width=\linewidth]{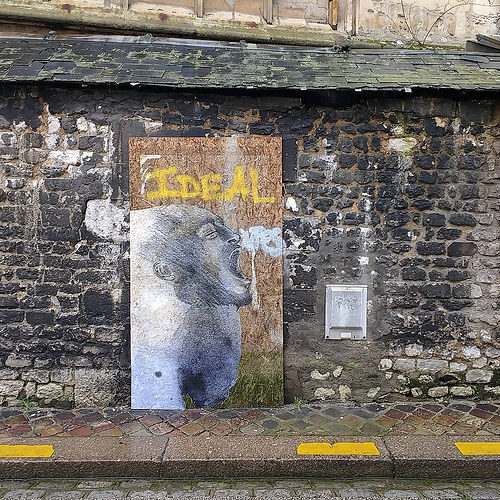}
    \caption{}
    \label{fig:case-3}
  \end{subfigure}
  \hfill
  \begin{subfigure}{0.21\textwidth}
    \includegraphics[width=\linewidth]{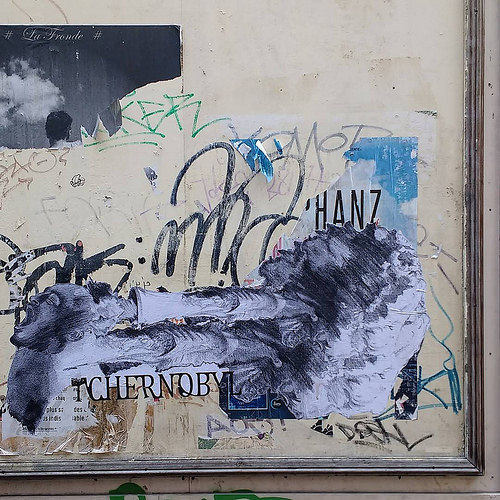}
    \caption{}
    \label{fig:case-4}
  \end{subfigure}
  \hfill
  \caption{Examples of posts where model predictions fail.}
  \label{fig:badcase}
\end{figure}

\section{Model Sensitivity Analysis}\label{app:param}
We analyze the sensitivity of several hyperparameters. In Figure ~\ref{fig:param-temp}, the model achieves the best performance when $\tau=0.1$, too small or too large a temperature parameter affects the model's ability to discriminate between samples. In Figure ~\ref{fig:param-cluster}, the number of clusters that achieve the best results is $40$, which is enough to distinct different user influence levels. Figure ~\ref{fig:param-drop} then shows the effect of the dropout ratio. The results are more stable when the ratio is small. This is because too large a dropout can impair the semantics of the features, resulting in augmented sample pairs that are no longer semantically similar. Figure ~\ref{fig:param-loss} demonstrates the effect of loss weight $\lambda$, and it can be seen that the higher the weight of the regression task, the effect is consistently improved. This suggests that the contrastive learning tasks only aid in adjusting the encoder and should not affect the supervised signal of the main regression task.
For the dimension of bottleneck layer in Post Encoder $d_b$, as shown in Figure \ref{fig:param-neck}, models on three datasets achieve the best results when $d_b=64$, suggesting that a relatively large $d_b$ can help the adapters learn new information from the downstream dataset. For the number of dense feature cross-layers in User Encoder $l$, from Figure \ref{fig:param-cross}, we can see the model performs well when $l=4$, too small a number of layers results in the user encoder not being able to adequately learn the associations between different numerical features, while too large a number of layers may result in the model overfitting the correlations between certain features.\par
Furthermore, since our model uses CLIP, a powerful multimodal pre-trained model, we test the model's performance under the use of lighter unimodal feature extractors. As shown in Table \ref{tab:vtfeat}, we replace CLIP-image and CLIP-text extractors\footnote{https://huggingface.co/openai/clip-vit-base-patch32} with ResNet-18\footnote{https://huggingface.co/microsoft/resnet-18} and BERT-base\footnote{https://huggingface.co/bert-base-uncased} (both are freezed), respectively. The model shows a little decrease in effectiveness, but still remains competitive compared to the baselines in Table ~\ref{tab:comparison}, effectively demonstrating the ability of our model to perform well despite weaker feature extractors, revealing the potential of modeling the impact of social factors on popularity prediction. 

Due to the recent boom in multimodal large language models (MLLMs)~\cite{liu2024improved,li2023blip2}, we also choose LLava-1.5 7b\footnote{https://github.com/haotian-liu/LLaVA} as a comparison. The prompt is the same as used for CLIP. Since the text branch used by Llava is a decoder-only model, we used the hidden state of the last token as the text feature. The dimensions of both visual and text features of CLIP we used are 512, while those of LLava are 1024 and 4096, respectively. Despite the much larger dimensions of the features, the results in Table ~\ref{tab:comparison} show that using LLava does not yield better results. We speculate that this may be caused by the following reasons: (1) The text data in SMPD is descriptions for images and is mostly simple, so the parameter number of CLIP may be sufficient to learn the post content information, and the use of overly complex pre-training models will do little to improve the results. (2) Since LLava's text branch is a decoder-only architecture, it is designed to specifically predict the next token, which is different from the encoding model. Therefore special handling may be required when using models like LLava for sentence encoding, such as the selection of token representations and constructing more appropriate prompts. Using MLLMs for popularity prediction is an interesting and promising direction, and we will leave more work for future attempts.


\section{case study}\label{app:case}
We use case studies to visually demonstrate the model’s predictions. As shown in Figure ~\ref{fig:case}, the upper right and lower left corners represent posts with successful predictions, while the other two quadrants represent posts with failed predictions. For the posts in the upper right corner, most of them have attractive photos, but there are also some posts that do not look popular (such as a strange face), and our model can successfully predict their actual high popularity. For the post in the lower left corner, most of the pictures look unattractive, but there are also some landscape photos that seem likely to be popular, and our model can successfully predict their low popularity. The above results demonstrate that our model is able to make inferences based on factors other than the attractiveness of the post content itself.\par
Then, we deeply analyze the reasons why the model fails to predict the posts in the other two quadrants. Figures ~\ref{fig:case-1} and ~\ref{fig:case-2} are cases where the model predicts a post with low popularity but actually has high popularity. First of all, the pictures in these two posts are not attractive. For Figure ~\ref{fig:case-1}, this post is the only one published by the user it belongs to, and the number of followers of this user is only $171$, which is relatively low. For Figure ~\ref{fig:case-2}, the average popularity of the user's historical posts is $5.5$, which is much lower than the popularity of this post of 7.5; at the same time, the number of followers of this user is only $17$. To summarize, the contrastive learning scheme we have designed does not work well for first-time users or for highly popular posts by users with a history of less popular posts (cold starts and mutation points). Figures ~\ref{fig:case-3} and ~\ref{fig:case-4} are cases where the model predicts a post with high popularity but actually has low popularity. These two posts are published by the same user, and they are also the only two posts of this user. However, despite the low number of posts, the number of followers of this user is as high as $968$, making him a "highly influential" user. Therefore, when the number of followers does not match the actual influence of the user, the UISD task we designed will mislead the model to make wrong judgments. This also inspires us to find more robust ways to express user influence in future work.

\section{Limitations}
In this paper, we attempt to explore the possibilities of social media popularity prediction from a new perspective, i.e., social impacts. However, there are still some potential issues to be addressed at present: (1) Due to the contrastive learning tasks designed for different data sources and labels, it is challenging to devise a sampling algorithm that ensures an adequate number of positive and negative samples for all contrastive tasks. Designing a sampling algorithm for one task may impact the effectiveness of other tasks. (2) The unsupervised augmentation algorithm we designed relies on dropout masks, which may not be applicable to popularity prediction models that do not use dropout. Therefore, to enhance the generality of our method, it is necessary to devise more generally applicable data augmentation methods. We will attempt to address the above issues in our future work. (3) Our model relies on labels to conduct contrastive learning. Although we try to construct labels in the UISD task, as shown in the case study, this approach may misclassify a small number of samples. Therefore, we need to design more robust ways to construct labels from the dataset to avoid adding wrong priors to some samples.

\end{appendix}

\end{document}